\documentclass{hch}
\usepackage{graphicx}

\begin{document}
\title{2D Transonic Hydrodynamics in General Relativity}
\author{Vasily Beskin}
\thanks{This work is supported by Grant N 00--15--96594 by Russian
Foundation for Basic Researches}
\address{Lebedev Physical Institute, Moscow, Russia}
\authorsup{Vasily Beskin\inst[Lebedev Physical Institute, Moscow]{1}}
\runningtitle{Beskin: 2D Transonic Hydrodynamics}

\maketitle

\begin{abstract}
The goal of my lecture is to present the introduction
into the hydrodynamical version of the Grad-Shafranov
equation. Although not so well-known as the full MHD one,
it allows us to clarify the nontrivial structure of the
Grad-Shafranov approach as well as to discuss the
simplest version of the $3+1$-split language -- the
most convenient one for the description of the ideal
flows in the vicinity of a rotating black hole.
\end{abstract}

\section{Introduction}

Axisymmetric stationary flows in the vicinity of a central compact body
have been studied for a long time in connection with many
astrophysical sources. Accretion onto ordinary stars
and black holes~\cite{bhoyle, bondi},
axially symmetrical stellar (solar) wind~\cite{parker},
jets from young stellar objects~\cite{bbr},
outflow from axisymmetric magnetosphere of rotating neutron
stars~\cite{michel, bgi, mestel} -- all of them are flows of the considered
type.

Let me stress that the necessity of taking into account the effects of
General Relativity is not so obvious for many compact sources.
For instance, one can not exclude that the black hole plays
only a passive role in the jet formation process, and the effects of General
Relativity in this case may be inimportant for flow
description in the region of jet formation.
At the same time gravitational effects make, apparently, a noticeable
contribution to the determination of physical conditions in compact objects.
First, this is indicated by the hard spectra and the annihilation line
observed in galactic X-ray sources, which are believed to be
solar mass black holes. Such characteristics are never observed in the
X-ray sources which are firmly established to show accretion not onto
a black hole but onto a neutron star~\cite{sunyaev}.
Another indication comes from superluminal motion in quasars which
may be due to the relativistic electron-positron plasma flow ejected
along with the weakly relativistic jet~\cite{guy}. All this testifies in
favour of the existence of an additional mechanism for particle creation
and acceleration, for which the effects of General Relativity may be
of principal importance. So, it is undoubtedly interesting to consider the flow
structure in the most general conditions, i.e., in the presence of a rotating
black hole.

There are several reasons why I restrict my consideration to the pure
hydrodynamic flows.
First of all, the hydrodynamical version of the Grad-Shafranov equation
is not so popular as the full MHD one. On the other hand, it contains all the
features of the full MHD version in the simplest form. In particular, within
the hydrodynamic approach one can introduce the $3+1$-split language -- the
most convenient one for the description of the ideal flows in the vicinity of
a rotating black hole.

Thus, in this lecture the basic equations describing a steady axisymmetric
hydrodynamical flow in the vicinity of Kerr black hole are given.
Starting with the well-known set of equations describing the nonrelativistic
ideal flow~\cite{hey},
we will go step by step to more complicated cases up to the most
general one corresponding to the axisymmetric stationary flow in the Kerr
metric. Finally, several examples will be considered which demonstrate how the
approach under consideration can be used to obtain the quantitative
description of the real transonic flows in the vicinity of rotating black
holes.

\section{Ideal Hydrodynamics -- Some Fundamental Results}

\subsection{Basic Equations}

To start from the very beginning, let us wright down the equations of
ideal stationary ($\partial/\partial t = 0$) hydrodynamics in a flat
space. They are~\cite{llhydro}:

\begin{itemize}
\item continuity equation
\begin{equation}
\nabla \cdot (n{\bf v})=0,
\label{a1}
\end{equation}
\item Euler equation
\begin{equation}
({\bf v}\nabla){\bf v} = -\frac{\nabla P}{m_{\rm p}n}-\nabla\varphi_{\rm g},
\label{a2}
\end{equation}
\item
isentropy condition
\begin{equation}
{\bf v} \cdot \nabla s = 0,
\label{a3}
\end{equation}
\item
equation of state
\begin{equation}
P=P(n,s).
\label{a4}
\end{equation}
\end{itemize}
The last expression can be rewritten in the form
\begin{equation}
{\rm d}P=m_{\rm p}n{\rm d} w-nT{\rm d}s.
\label{a5}
\end{equation}
Here $n$ (${\rm cm}^{-3}$) is the concentration,
$s$ is the entropy per particle (undimensional),
$w$ (${\rm cm}^2/{\rm s}^2$) is the specific enthalpy,
$m_{\rm p}$ (g) is the mass of particle,
$T $ (erg) is the temperature in energetic unit,
and $c_s $ (cm/s) is the velocity of sound.
For the polytropic equation of state
\begin{equation}
P=k(s)n^{\Gamma},
\label{a6}
\end{equation}
used for simplicity in what follows, one can obtain for $\Gamma \neq 1$
\begin{eqnarray}
c_s^2 & = & \frac{1}{m_{\rm p}}
\left(\frac{\partial P}{\partial n}\right)_s
= \frac{1}{m_{\rm p}}\Gamma k(s)n^{\Gamma-1},
\label{a7}   \\
w & = & \frac{c_s^2}{\Gamma-1},
\label{a8} \\
T & = & \frac{m_{\rm p}}{\Gamma}c_s^2.
\label{a9}
\end{eqnarray}

Now one can make several remarks.

\begin{itemize}
\item
The Euler equation (\ref{a2}) together with (\ref{a3}) and (\ref{a5}) can be
rewritten as the energy equation
\begin{equation}
\nabla \cdot
\left[n{\bf v}\left(\frac{v^2}{2} + w + \varphi_{\rm g}\right)\right] = 0.
\label{a10}
\end{equation}
Now using the continuity equation (\ref{a1}), one can obtain
\begin{equation}
{\bf v} \cdot \nabla E = 0,
\label{a10a}
\end{equation}
where
\begin{equation}
E = \frac{v^2}{2} + w + \varphi_{\rm g}.
\label{a10b}
\end{equation}
This is the well-known Bernoulli integral.
\item
The energy equation (\ref{a10}) together with the Euler equation (\ref{a2})
can be rewritten as a four-component energy-momentum equation
\begin{equation}
\nabla_{\alpha}T^{\alpha\beta}=0,
\label{a11}
\end{equation}
where for $\varphi_{\rm g} = 0$
\begin{equation}
T^{\alpha\beta} = \pmatrix{nm_{\rm p}v^2/2 + n\varepsilon
&& nm_{\rm p}{\bf v}\left(v^2/2+w\right)\cr
&&\cr
m_{\rm p}nv^i
&&P\delta^{ik}+nm_{\rm p}v^iv^k\cr}.
\label{a12}
\end{equation}
Here and below the Greek indices $\alpha$ and $\beta$ correspond
to four-di\-men\-sion\-al values, while the Latin $i$, $j$, and $k$ are
three-dimensional.
\item
As a result, hydrodynamics contains five equations over five unknown values.
\end{itemize}

\vspace*{.5cm}

{\bf exercise}
\begin{enumerate}
\item
{\it Prove expressions (\ref{a10})--(\ref{a12})}
\end{enumerate}

\subsection{An Example -- Spherically Symmetrical Flow}

As a most simple but very important example let us
consider the spherically symmetrical flow.
Since the basic (ideal) hydrodynamic equations have
the form of the conservation laws, one can find
\begin{itemize}
\item
the continuity equation
\begin{equation}
\Phi = 2\pi r^2n(r)v(r) = {\rm const},
\label{a13}
\end{equation}
\item
the entropy equation
\begin{equation}
s = {\rm const},
\label{a14}
\end{equation}
\item
the energy equation
\begin{equation}
E=\frac{v^2(r)}{2}+w(r)+\varphi_{\rm g}(r) = {\rm const}.
\label{a15}
\end{equation}
\end{itemize}

As a result, knowing three parameters $\Phi$, $s$, and $E$, one can
determine all the physical characteristics of a flow. Indeed, rewriting
the Bernoulli equation (\ref{a15}) as
\begin{equation}
E=\frac{\Phi^2}{8\pi^2n^2r^4}+w(n,s)+\varphi_{\rm g}(r),
\label{a16}
\end{equation}
we see that this equation contains only one unknown
parameter $n$. Hence, this algebraic equation implicitly determines
the concentration $n$ as a function of the invariants and radius $r$:
\begin{equation}
n=n(E,s,\Phi;r).
\label{a17}
\end{equation}
Together with the entropy $s$, it allows us to determine all the other
thermodynamic functions and the flow velocity $v(r)$.

It is necessary to stress that equation (\ref{a16}) contains a
singularity on the sonic surface. To show this, let us determine the
derivative ${\rm d}n/{\rm d}r$:
$$
\frac{{\rm d}n}{{\rm d}r}\left[\left(\frac{\partial w}{\partial n}\right)_s
-\frac{\Phi^2}{4\pi^2n^3r^4}\right]
-\frac{\Phi^2}{2\pi^2n^2r^5}+\frac{GM}{r^2}=0.
$$
As a result, using the thermodynamic relation (\ref{a5}),
one can obtain for the logarithmic derivative
\begin{equation}
\eta_1 = \frac{r}{n}\frac{{\rm d}n}{{\rm d}r}=
\frac{2v^2-GM/r}{c_s^2-v^2}= \frac{2-G{\cal
}M/rv^2}{-1+c_s^2/v^2}=\frac{N_r}{D}.
\label{a18}
\end{equation}
We see that the derivative (\ref{a18}) contains the singularity when the
velocity is equal to that of sound $c_{*}$ ($D=0$). It means that to pass
through the sonic surface $r=r_{*}$ the additional critical condition
is to be valid:
\begin{equation}
N_r(r_{*})= 2 - \frac{GM}{r_{*}c_{*}^2} = 0.
\label{a19}
\end{equation}
In other words, the transonic flows are determined by
two invariants only. As shown in Fig. 1, the sonic surface is an X-point on
the plane distance -- velocity.

\begin{figure}[hbtp]
        \centering
        \includegraphics[width=9.truecm]{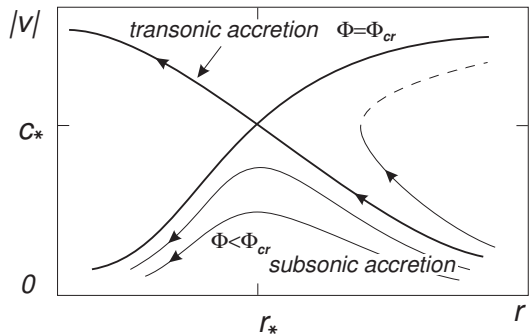}
        \caption[]{Structure of the spherically symmetric accretion
         for a given $n_{\infty}$ and $c_{\infty}$, and different
         stream function $\Phi$. The transonic flow corresponds to
         the critical accretion rate $\Phi = \Phi_{\rm cr}$
         (\ref{rate}). The curves below the X-point correspond
         to subsonic accretion with $\Phi < \Phi_{\rm cr}$}
        \label{fig1}
\end{figure}

\vspace*{.5cm}

{\bf exercises}

{\it

\begin{enumerate}
\item
For spherically symmetrical transonic inflow (Bondi accretion~\cite{bondi})
one can determine the Bernoulli integral $E$
through the velocity of sound at infinity
$$
E= w_{\infty}=\frac{c_{\infty}^2}{\Gamma-1}.
$$
Now using relations (\ref{a13})--(\ref{a15}) and (\ref{a19}),
obtain the well-known expressions
for the velocity of sound $c_{*}$ and the concentration $n_{*}$ on the
sonic radius $r_{*}$~\cite{bondi}:
\begin{eqnarray}
c_{*}^2 & = & \left(\frac{2}{5-3\Gamma}\right)c_{\infty}^2,
\label{a20} \\
n_{*} & = & \left(\frac{2}{5-3\Gamma}\right)^{1/(\Gamma-1)}n_{\infty},
\label{a21} \\
r_{*} & = & \left(\frac{5-3\Gamma}{4}\right)\frac{GM}{c_{\infty}^2}.
\label{a22}
\end{eqnarray}
\item
Show that
\begin{equation}
\eta_1(r_{*})=\frac{-4\pm\sqrt{10-6\Gamma}}{\Gamma+1},
\label{a23}
\end{equation}
the sign {\rm plus} corresponding to accretion and {\rm minus} -- to
ejection.
\item
Find that for the spherically symmetric accretion
\begin{itemize}
\item
for $r\gg r_{*}$ (subsonic regime) the flow is approximately
incompressible:
\begin{eqnarray}
n(r) & \approx &  const,
\label{a24} \\
v(r) & \propto & r^{-2}.
\label{a25}
\end{eqnarray}
\item for $r\ll r_{*}$ (supersonic regime) the particle motion corresponds
to a free fall:
\begin{eqnarray}
n(r) & \propto &  r^{-3/2},
\label{a26} \\
v(r) & \approx & (2GM/r)^{1/2}.
\label{a27}
\end{eqnarray}
\end{itemize}
\item
Find that for the spherically symmetric transonic outflow
(Parker ejection~\cite{parker}):
\begin{itemize}
\item
Physical parameters on the sound surface $r = r_{*}$, where again
\begin{equation}
r_{*} = \frac{GM}{2c_{*}^2},
\label{p2}
\end{equation}
have the following relations to the ones on the star surface $r = R$
\begin{eqnarray}
c_{*}^2 & = & \left(\frac{2}{5-3\Gamma}\right)c_{R}^2
+ \left(\frac{\Gamma-1}{5-3\Gamma}\right)\left(v_{R}^2-\frac{2GM}{R}\right),
\label{p1} \\
n_{*} & = & n_{R}\left(\frac{c_{*}^2}{c_{R}^2}\right)^{1/(\Gamma-1)}.
\label{p3}
\end{eqnarray}
\item
The (radial) velocity on the star surface is to be
\begin{equation}
v_{R}=c_{*}\left(\frac{c_{*}^2}{c_{R}^2}\right)^{1/(\Gamma-1)}
\left(\frac{r_{*}}{R}\right)^2.
\end{equation}
\end{itemize}
\end{enumerate}

}

Although fairly simple, the radial 1D flow allows us
to formulate here several important properties; some of them, as we shall
see, retains the common properties of the Grad-Shafranov approach.

\begin{itemize}
\item
It is possible to pass through the sonic surface in the presence
of gravity only. Indeed, the nominator $N_r$ in (\ref{a18}) can vanish
for $D = 0$ in the presence of the gravity term $GM/rv^2$ only.
\item
Solutions (\ref{a20})--(\ref{a22}) and (\ref{p1})
have singularity for $\Gamma=5/3$. It means that for
$\Gamma=5/3$ the increase/decrease of the velocity of sound as a
result of adiabatic heating/cooling equals the change of particle
velocity. As a result, in the nonrelativistic case
for $\Gamma \geq 5/3$ the transonic flow is not
realized.
\item
The transonic problem is two-parametric. It means that to determine the
transonic flow it is necessary to specify two boundary conditions, say,
the density $m_{\rm p}n_{\infty}$ and the velocity of sound $c_{\infty}$ at
infinity. In particular, the accretion rate is fixed:
\begin{equation}
2\Phi_{\rm cr} = 4\pi r_{*}^2c_{*}n_{*}
= 4\pi\left(\frac{2}{5-3\Gamma}\right)^{(5-3\Gamma)/2(\Gamma+1)}
\frac{(GM)^2}{c_{\infty}^3}n_{\infty}.
\label{rate}
\end{equation}
On the other hand, for a given $n_{\infty}$ and $c_{\infty}$
there is an infinite number of subsonic solutions
with $\Phi < \Phi_{\rm cr}$ (see Fig. 1).
\item
For a given flow structure the number of integrals is enough to determine
all the characteristics of a flow from algebraic equations.
\end{itemize}

The last property is actually the key point of the approach.
Indeed, the algebraic relations (\ref{a13})--(\ref{a15}) together with
the equation of state allow us to determine all the physical parameters
of the flow (the flow velocity $v(r)$, the temperature $T(r)$,
{\it etc}.) through the invariants $E$ and $s$ and the stream function $\Phi$.
This property remains true for an arbitrary 2D flow structure. But in the
general case the flow structure (i.e., the stream function $\Phi(r,\theta)$)
itself is not known. To determine the stream function, the extra two
hydrodynamic equations are to be used.

\subsection{Flat Potential Flow}

To start, let us consider the simplest (and the well-known)
case of the flat potential flow without gravity. Then one
can introduce potential $\phi(x,y)$
by the relation
\begin{equation}
{\bf v}=\nabla\phi(x,y).
\label{a30}
\end{equation}
In addition, let us consider the case
\begin{equation}
E={\rm const}, \qquad s={\rm const}.
\end{equation}
Then the continuity equation $\nabla \cdot (n{\bf v})=0$ can be rewritten as
\begin{equation}
\nabla^2\phi+\frac{\nabla n\cdot\nabla\phi}{n} = 0.
\label{a31}
\end{equation}
Finally, using the Euler equation to determine
$\nabla n\cdot\nabla\phi$
$$
{\bf v} \cdot \nabla\left(\frac{v^2}{2}\right)
+c_{\rm s}^2\frac{\nabla n\cdot\nabla\phi}{n} = 0,
$$
we obtain
\begin{equation}
\phi_{xx}+ \phi_{yy}+\frac{ (\phi_y)^2\phi_{xx}
-2\phi_x\phi_y\phi_{xy} +(\phi_x)^2\phi_{yy}}
{(\nabla\phi)^2D}=0.
\label{phi}
\end{equation}
Here again
\begin{equation}
D=-1+\frac{c_s^2}{v^2}.
\label{phia}
\end{equation}
This well-known equation can be found in any textbook
(e.g., see \cite{llhydro}).

For us it is necessary to stress here the following properties.
\begin{itemize}
\item
To determine $c_s^2$, equation (\ref{phi}) is to be supplemented with the
Bernoulli equation. For the polytropic equation of state it can be solved for
the velocity of sound $c_s$:
\begin{equation}
c_s^2=(\Gamma-1)E-\frac{\Gamma-1}{2}(\nabla\phi)^2.
\label{a33}
\end{equation}
\item
Together with the Bernoulli equation, equation (\ref{phi}) contains the
potential $\phi(x,y)$ and the invariant $E$ only (it does not contain the
entropy $s$ at all, but $s$ is necessary to determine $n$).
\item
For $n = {\rm const}$ ($c_s^2 \rightarrow \infty$) the equation becomes linear.
\item
It is nonlinear in the general case, but linear for second derivatives.
\item
The equation is elliptical for a subsonic flow $D > 0$.
\item
The equation is hyperbolic for a supersonic flow $D < 0$.
\item
For a given flow structure (for a given $\phi$, $E$, and $s$) all the physical
parameters are determined by algebraic relations.
\item
Equation (\ref{phi}) does not contain coordinates $x$ and $y$.
\end{itemize}

The latter property was very widely used in the approach of the
hodograph transformation, i.e., the transformation from a physical plane
$(x,y)$ to a hodograph plane $(v,\theta)$, where
$v_x=v\sin\theta$, $v_y=v\cos\theta$.
In this case it is possible to introduce another potential
$\phi_v(v,\theta)$ so that ${\bf r}=\nabla_{\bf v}\phi_v$.
As a result, equation (\ref{phi}) can be rewritten as
\begin{equation}
\frac{\partial^2\phi_v}{\partial\theta^2}
+\frac{v^2}{1-v^2/c_s^2}\frac{\partial^2\phi_v}{\partial v^2}
+v\frac{\partial\phi_v}{\partial v}=0.
\label{a34}
\end{equation}
This is the linear Chaplygin equation (1902).

The hodograph transformation approach was the main direction of exploration
through the XX-th century~\cite{vonmises, guderley}.
Here I formulate two results
obtained in this field, to be used in what follows.

\begin{itemize}
\item
For the transonic flow it is impossible to solve the direct problem
(i.e., to determine the flow structure from the given shape of the boundary,
say, knowing the shape of nozzle or wing).
\item
On the other hand, one can solve the reverse problem. This approach is based
on the fundamental theorem: the transonic flow is analytical at the critical
point (the point where the sonic surface is orthogonal to the flow
line, see Fig. 2)~\cite{llhydro}.
\end{itemize}

Let me comment these two statements. The most visible argument clarifying
the absence of the regular (not iterative) procedure for the transonic flow
is as follows. As is known,
the number of boundary conditions $b$ for an arbitrary (not only for a
pure hydrodynamical) flow can be determined from the following
condition~\cite{ufn, bogovalov}
\begin{equation}
b = 2 + i - \sigma.
\label{a35}
\end{equation}
Here $i$ is the number of invariants and $\sigma$ is the number of singular
surfaces. In pure hydrodynamics the only singular surface is the sonic one.
Hence, for the transonic flow $\sigma = 1$. Furthermore, for planar geometry
we have two invariants, $E$ and $s$, so that $i = 2$. Thus, to determine
the structure of the transonic flow it is necessary to
specify three boundary conditions on a surface. These may be two thermodynamic
functions and one component of the velocity. The second (the last in the planar
case) component of the velocity is to be determined from the solution. But to
solve equation (\ref{phi}), it is necessary to know the Bernoulli integral
$E = v^2/2 + w$, i.e., both components of the velocity on this surface.
Hence, in the general case even the equation describing the flow structure
cannot be formulated. For subsonic and supersonic flows $\sigma = 0$
(so that $b = 4$) and this difficulty is absent.

\begin{figure}[hbtp]
        \centering
        \includegraphics[width=9.truecm]{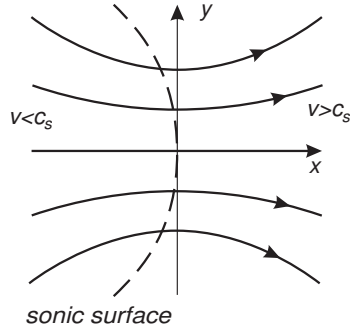}
        \caption[]{Structure of the "analytical nozzle"
        in the vicinity of the critical point $x = y = 0$
        -- the only point where the sonic surface is orthogonal
        to the flow line. As the term $\propto xy$ in (\ref{a37})
        is absent (i.e., $v_y(0,y) = 0$), the plane $x = 0$
        corresponds to minimum cross-section of the stream
        surfaces. The sonic surface $v = c_{*}$ has the standard
        parabolic form $x = -[k(\Gamma - 1)/2c_*]y^2$}
        \label{fig2}
\end{figure}

On the other hand, the structure of the transonic flow can be found
directly by expansion of the solution in the vicinity of the critical
point (where we put $x = y = 0$). Indeed, in addition to the invariants
$E$ (which determines the velocity of sound on the sonic surface) and
$s$ (which is necessary for the determination of the concentration $n$)
one can specify the $x$ component of the velocity $v_x(x,0)$ along $x$-axis.
For our purpose it is enough to know the first two terms in the expansion
\begin{equation}
v_x(x,0) = c_* + kx + \dots
\label{a36}
\end{equation}
Here $c_*^2 = 2E(\Gamma-1)/(\Gamma+1)$.
As a result, as one can check directly, the first terms in the expansion
of the potential $\phi(x,y)$ look like~\cite{llhydro}
\begin{equation}
\phi(x,y) = c_*x +\frac{1}{2}\,kx^2 + \frac{1}{2c_*}\,k^2(\Gamma+1) xy^2
+ \frac{1}{24c_*^2}\,k^3(\Gamma+1)^2y^4 + \dots
\label{a37}
\end{equation}
Knowing all the coefficients in the expansion
(\ref{a36}), one can restore the potential $\phi$ with any
precision.

Thus, in the general case equation (\ref{phi}) cannot be solved directly.
Let me stress that this property is common; it takes place
for the Grad-Shafranov equation as well.
On the other hand, the planar approach has
three extra difficulties:

\begin{itemize}
\item
It is difficult to consider the case $E \neq$ const,
$s \neq$ const.
\item
It is impossible to consider the nonpotential flow with
$\nabla \times {\bf v} \neq 0$.
\item
It is impossible to include gravity (which is not planar).
\end{itemize}

\section{Axisymmetric Stationary Flow -- Nonrelativistic Case}

\subsection{Basic Equations}

\subsubsection{Stream Function}

Now let us see how a similar procedure can be realized for the axisymmetric
stationary flows. It means that we assume all the values to
depend on two variables -- $r$ and $\theta$.
In this sense the flow remains two-dimensional. But now none of the three
components of the velocity are equal to zero.
Thus, axisymmetric stationary flows are more rich than planar ones.

In the axisymmetric stationary case one can introduce potential
$\Phi(r,\theta)$ which is connected with the poloidal velocity
${\bf v}_{\rm p}$ as
\begin{equation}
n{\bf v}_{\rm p}
= \frac{\nabla \Phi \times {\bf e}_{\varphi}}{2\pi r\sin\theta}.
\label{b1}
\end{equation}
Such a definition results in the following properties:

\begin{itemize}
\item
The continuity equation is valid automatically:
$\nabla \cdot (n {\bf v}) = 0$.
\item
On can check that ${\rm d} \Phi = n{\bf v} {\rm d} {\bf S}$, where ${\bf S}$
is the area. Thus, the potential $\Phi(r,\theta)$ is the flux
through the circuit $r, \theta, 0<\varphi<2\pi$.
In particular, the total flux through the sphere of radius $r$ is
$\Phi_{\rm tot} = \Phi(r,\pi)$.
\item
As ${\bf v} \cdot \nabla \Phi = 0$, the relations $\Phi(r,\theta) = $ const
describe the flow surfaces.
\end{itemize}

\subsubsection{Integrals of motion}

As earlier, the components $\beta = t$ and $\beta = v_{\parallel}$ of the
energy-momentum conservation law $\nabla_{\alpha} T^{\alpha\beta} = 0$
give
\begin{eqnarray}
E & = & E(\Phi) = v^2/2 + w + \varphi_{\rm g},
\label{b2} \\
s & = & s(\Phi).
\label{b3}
\end{eqnarray}
But as we see, it is now much easier to describe the case when integrals are
different for different flow surfaces.

New information appears from the $\beta = \varphi$ component of the
energy-momentum equation (or, which is the same, from the $\varphi$
component of the Euler equation)
\begin{equation}
\nabla_{\varphi}\left(\frac{v^2}{2}\right)
- [{\bf v} \times(\nabla\times{\bf v})]_{\varphi}
+ \frac{\nabla_{\varphi} P}{m_{\rm p}n}
+ \nabla_{\varphi} \varphi_{\rm g} = 0.
\label{b4}
\end{equation}
As we consider the axisymmetric case, all the gradients vanish.
The last term
$[{\bf v} \times(\nabla\times{\bf v})]_{\varphi}$
can be rewritten in the form of the conservation law
\begin{equation}
{\bf v} \cdot \nabla (r\sin\theta v_{\varphi}) = 0.
\label{b5}
\end{equation}
Hence, in the axisymmetric case the $z$-component of the angular momentum
\begin{equation}
L(\Phi) = v_{\varphi}r\sin\theta
\label{b6}
\end{equation}
is the third integral of motion.

\subsubsection{Mathematical Interlude -- The Covariant Description}

As we are going to generalize our equations to General Relativity,
it is convenient to rewrite these relations right now in the covariant
form. For this reason, recall that the flat 3D metric $g_{ik}$
(${\rm d}s^2 = g_{ik}{\rm d}x^i{\rm d}x^k$) for spherical coordinates
$x^1 = r$, $x^2 = \theta$, and $x^3 = \varphi$ has a form
\begin{equation}
g_{rr} = 1, \qquad g_{\theta\theta} = r^2,
\qquad g_{\varphi\varphi} = r^2\sin^2\theta,
\label{metricp}
\end{equation}
all the other components being zero.
Using now expression
(\ref{a12})
\begin{equation}
T_i^k = P \delta_i^k + (n m_{\rm p})v^kv_i,
\label{b7}
\end{equation}
one can obtain from the $\varphi$ component of the energy-momentum equation
\begin{eqnarray}
 & & \nabla_k T_{\varphi}^k = \nabla_{k}(\delta^k_{\varphi}P)
+ \nabla_k (n m_{\rm p}v^kv_{\varphi})
= \frac{1}{r}\,\frac{\partial P}{\partial \varphi}
\nonumber \\
 & & + \frac{\partial}{\partial x^k}(nm_{\rm p}v^kv_{\varphi})
+ \Gamma^k_{ik}(nm_{\rm p})v^iv_{\varphi}
- \Gamma^k_{\varphi i}(nm_{\rm p})v^iv_k = 0.
\label{b8}
\end{eqnarray}
Here $\Gamma^{i}_{jk}$ are the Christoffel coefficients.

As one can check, the last term in (\ref{b8}) vanishes:
$\Gamma^k_{\varphi i}(nm_{\rm p})v^iv_k = 0$.
The first term vanishes because of axisymmetry of the problem
(no $\varphi$ -- dependence). Using now the continuity equation
\begin{equation}
\nabla_k(nm_{\rm p}v^k) = \frac{1}{\sqrt{g}}\frac{\partial}{\partial x^k}
(\sqrt{g}nm_{\rm p}v^k) =
\frac{\partial}{\partial x^k}(nm_{\rm p}v^k)
+\Gamma^k_{ik}nm_{\rm p}v^i = 0,
\label{b9}
\end{equation}
where $g = \det g_{ik} = g_{rr}g_{\theta\theta}g_{\varphi\varphi}$,
we see that (\ref{b8}) can again be written down in the form of the
conservation law
$\nabla_k T_{\varphi}^k = (n m_{\rm p}){\bf v} \cdot \nabla v_{\varphi}$.
Hence, the third invariant looks like
\begin{equation}
L(\Phi)  = v_{\varphi}.
\label{b10}
\end{equation}

\vspace{.3cm}

{\bf exercises}

{\it

\begin{enumerate}
\item
Check Eqns. (\ref{b8})--(\ref{b10}).
\item
Is there a contradiction between (\ref{b6}) and (\ref{b10})?
\end{enumerate}
}

\vspace{.3cm}

To understand the difference between (\ref{b6}) and (\ref{b10}),
it is necessary to return to the main definitions of the covariant approach.
Up to now we have dealt with the physical components only.
Below in the relativistic expressions in Sect. 4
we will mark the physical components by hats, so that
$v_{\hat\varphi} = v^{\hat\varphi}$ is the physical component of the toroidal
velocity, and its dimension is cm/s. But in the covariant expressions
(\ref{b8})--(\ref{b10}) we encounter other objects -- contravariant
components $v^i$ and covariant components $v_k$. As the definition of the
vector length (which is square of the physical component)
has the form ${\bf v}^2 = g_{ik}v^iv^k = g^{ik}v_iv_k$, one can write down
for the diagonal metric (\ref{metricp})
\begin{equation}
(v_{\hat\varphi})^2 = g_{\varphi\varphi}(v^{\varphi})^2 =
g^{\varphi\varphi}(v_{\varphi})^2,
\label{b11}
\end{equation}
and the same for the other components. Thus,
\begin{eqnarray}
v^{\varphi} & = & \frac{1}{\sqrt{g_{\varphi\varphi}}}v_{\hat\varphi},
\label{b12} \\
v_{\varphi} & = & \sqrt{g_{\varphi\varphi}}v_{\hat\varphi}.
\label{b13}
\end{eqnarray}
In particular, it means that the dimension of the covariant and contravariant
components may differ from the dimension of the physical one. Comparing
now (\ref{b13}) with (\ref{b6}) and (\ref{b10}), we can understand the
difference: (\ref{b6}) actually involves the physical component
of the toroidal velocity while (\ref{b10}) includes the covariant one.

\subsection{The Stream Equation}

\subsubsection{Grad-Shafranov Equation}

To formulate the stream equation (i.e., the equation describing the stream
function $\Phi(r,\theta)$) it is necessary to return to the poloidal
component of the Euler equation. One can check that together with
the definitions of the invariants $E(\Phi)$, $L(\Phi)$, and $s(\Phi)$
this vector equation can be written as the product of the
scalar equation by the vector $\nabla\Phi$:
$[{\rm Euler}]_{\rm p} = [{\rm GS}] \cdot \nabla\Phi$.  For this reason,
in many MHD papers the stream equation $[{\rm GS}] = 0$ was obtained as the
projection of the poloidal equation onto the gradient $\nabla\Phi$. The
equivalent hydrodynamic expression has the form
\begin{equation}
\frac{1}{(\nabla\Phi)^2}\nabla\Phi \cdot
\left[({\bf v}\nabla){\bf v} + \frac{\nabla P}{nm_{\rm p}}
+ \nabla\varphi_{\rm g}\right] = 0.
\label{b14}
\end{equation}
Using now the definitions (\ref{b1}), (\ref{b2}), and (\ref{b6}),
we have
\begin{eqnarray}
 & & -\varpi^2\nabla_k\left(\frac{1}{\varpi^2}\nabla^k\Phi\right)
+\frac{1}{n}\nabla_kn\cdot\nabla^k\Phi
\nonumber \\
\label{b15}
& & -4\pi^2L\frac{{\rm d}L}{{\rm d}\Phi}
+4\pi^2\varpi^2n^2\frac{{\rm d}E}{{\rm d}\Phi}
-4\pi^2\varpi^2n^2\frac{T}{m_{\rm p}}\frac{{\rm d}s}{{\rm d}\Phi} = 0.
\end{eqnarray}
Here and to the very end
\begin{equation}
\varpi = \sqrt{g_{\varphi\varphi}},
\label{b16}
\end{equation}
so that for the flat metric $\varpi = r\sin\theta$.

To close the system, i.e., to determine the product
($\nabla n \cdot \nabla\Phi$), the stream equation (\ref{b15})
is to be supplemented with the Bernoulli equation (\ref{b2}).
It can now be rewritten in the form (cf. (\ref{a16}))
\begin{equation}
E = \frac{(\nabla\Phi)^2}{8\pi^2\varpi^2n^2}
+ \frac{1}{2}\,\frac{L^2}{\varpi^2} + w(n,s) + \varphi_{\rm g}.
\label{b17}
\end{equation}
As previously, the Bernoulli equation (\ref{b17}) contains, besides $n$,
the invariants $E$, $L$, and $s$, and the stream function $\Phi$
only. Hence, it again gives the implicit expression for $n$:
\begin{equation}
n = n(\nabla\Phi;E,L,s;r,\theta).
\label{b18'}
\end{equation}
On the other hand, the implicit Bernoulli equation can be presented
in the differential form
\begin{equation}
\nabla_k n = n\frac{N_k}{D},
\label{b18}
\end{equation}
where now
\begin{equation}
D = -1 +\frac{c_{\rm s}^2}{v_{\rm p}^2},
\label{b19}
\end{equation}
and
\begin{eqnarray}
 & & N_k = -\frac{\nabla^{i}\Phi \cdot \nabla_{i}\nabla_{k}\Phi}{(\nabla
\Phi)^2} + \frac{1}{2}\cdot\frac{\nabla_{k}\varpi^2}{\varpi^2}-
4\pi^{2}\varpi^{2}n^2\frac{\nabla_{k}\varphi_{\rm g}}{(\nabla\Phi)^2}
\nonumber \\
 & & - 4\pi^2n^2L\frac{{\rm d}L}{{\rm d}\Phi}
\frac{\nabla_{k}\Phi}{(\nabla\Phi)^2}
+ 2\pi^2n^2L^2\frac{\nabla_{k}\varpi^2}{\varpi^2(\nabla\Phi)^2}
\label{b20}\\
 & &+ 4\pi^{2}\varpi^{2}n^2
\frac{{\rm d}E}{{\rm d}\Phi}
\frac{\nabla_{k}\Phi}{(\nabla\Phi)^2}
- 4\pi^2\varpi^{2}n^2\left[\frac{T}{m_{\rm p}}
+\frac{1}{m_{\rm p}n}\left(\frac{\partial P}{\partial s}\right)_{n}\right]
\frac{{\rm d}s}{{\rm d}\Phi}
\frac{\nabla_{k}\Phi}{(\nabla\Phi)^2}.
 \nonumber
\end{eqnarray}
As a result, the stream equation can be written down as~\cite{ufn}

\begin{eqnarray}
 &  & -\varpi^2\nabla_{k}
\left(\frac{1}{\varpi^{2}}\nabla^{k}\Phi\right)
-\frac{\nabla^{i}\Phi\cdot\nabla^{k}\Phi\cdot\nabla_{i}
\nabla_{k}\Phi}{(\nabla\Phi)^2 D}
+\frac{\nabla\varpi^2\cdot\nabla\Phi}{2D\varpi^2}
\nonumber  \\
 &  & - 4\pi^2\varpi^{2}n^2\frac{\nabla\varphi_{\rm g}\cdot\nabla\Phi}
{D(\nabla\Phi)^2}
-4\pi^{2}n^2\frac{D+1}{D}L\frac{{\rm d}L}{{\rm d}\Phi}
\label{gs} \\
 &  & +2\pi^{2}n^2\frac{\nabla(\varpi^2)\cdot\nabla\Phi}
{D\varpi^2(\nabla\Phi)^2}L^2
+4\pi^{2}\varpi^{2}n^2\frac{D+1}{D}\frac{{\rm d}E}{{\rm d}\Phi}
\nonumber \\
& & -4\pi^2\varpi^{2}n^2
\left[\frac{D+1}{D}\frac{T}{m_{\rm p}}
+\frac{1}{Dm_{\rm p}n}\left(\frac{\partial P}{\partial s}
\right)_{n}\right]\frac{{\rm d}s}{{\rm d}\Phi}=0,
\nonumber
\end{eqnarray}
or, in a compact form (cf.~\cite{heyvaerts}), as

\begin{eqnarray}
 &  & -\varpi^2 \nabla_{k}
\left(\frac{1}{\varpi^{2}n}\nabla^{k}\Phi\right)
-4\pi^{2}nL\frac{{\rm d}L}{{\rm d}\Phi}
\label{gscomp} \\
 &  &
+4\pi^{2}\varpi^{2}n\frac{{\rm d}E}{{\rm d}\Phi}
-4\pi^2\varpi^{2}n\frac{T}{m_{\rm p}}\frac{{\rm d}s}{{\rm d}\Phi}=0.
\nonumber
\end{eqnarray}
At first glance, the stream equation (\ref{gs}) is much more complicated
than the planar one (\ref{phi}). Nevertheless, one can easily see that
these equations have many similarities.  As (\ref{phi}), the stream
equation (\ref{gs}) starts with the linear elliptic term and the nonlinear
term with an analogous form. The third term does not, of course, exist
in (\ref{phi}) -- it results from non-cartesian geometry. But all
the other terms are not complification. They allow us to include into
consideration not only gravity, but a much wider class of flows
with different invariants onto different flow surfaces.

In other respects the stream equation is quite similar to the planar equation
(\ref{phi}):
\begin{itemize}
\item
The stream equation (\ref{gs}) is to be supplemented with the
Bernoulli equation.
\item
Together with the Bernoulli equation, equation (\ref{gs}) contains the
potential $\Phi(r,\theta)$ and the invariants $E$, $L$, and $s$ only (i.e.,
it has the Grad-Shafranov form).
\item
For $n = {\rm const}$ ($c_s^2 \rightarrow \infty$), $E = $ const,
$s =$ const, and $L = 0$ the equation becomes linear.
\item
It is nonlinear in the general case, but linear for second derivatives.
\item
The equation is elliptical for a subsonic flow $D > 0$.
\item
The equation is hyperbolic for a supersonic flow $D < 0$.
\item
For a given flow structure (i.e., for a given $\Phi$) and for the
invariants $E(\Phi)$, $L(\Phi)$, and $s(\Phi)$ all the physical
parameters are determined by algebraic relations.
\end{itemize}

The following point should be stressed. The expression for the
denominator $D = -1 + c_{\rm s}^2/v_{\rm p}^2$ (\ref{b19})
involves the poloidal rather than
the total velocity. It means that the sonic surface exists
when the poloidal, not the total velocity becomes equal to that
of sound. This fact results from our basic assumption: as we consider
axisymmetric flows only, the disturbances (waves) are to have the same
symmetry as well. Hence, the disturbances can propagate in the poloidal
direction only. For this reason, a singularity appears at the moment
when the particle velocity coincides with the velocity of disturbance.

\subsubsection{Linear Operator $\varpi^2\nabla_k(\varpi^{-2}\nabla^k)$}

In what follows we shall use the linear operator
\begin{equation}
\hat{\cal L} = \varpi^2\nabla_k\left(\frac{1}{\varpi^{2}}\nabla^k\right) =
\frac{\partial^2}{\partial r^2}
+ \frac{\sin\theta}{r^2}\frac{\partial}{\partial \theta}
\left(\frac{1}{\sin\theta}\,\frac{\partial}{\partial\theta}\right)
\label{b21}
\end{equation}
and shall therefore consider it in more detail.
First, examine the angular operator
\begin{equation}
\hat{\cal L}_{\theta}=\sin\theta\frac{\partial}{\partial\theta}\left(\frac{1}{
\sin\theta} \frac{\partial}{\partial\theta}\right).
\label{b22}
\end{equation}
It has the following eigenfunctions
\begin{eqnarray}
  &  & Q_0=1-\cos\theta ,
\label{b23}  \\
  &  & Q_1=\sin^{2}\theta ,
\label{b24}  \\
  &  & Q_2=\sin^{2}\theta \cos\theta ,
\label{b25} \\
  &  &  \dots \nonumber  \\
  &  & Q_m=\frac{2^{m}m!(m-1)!}{(2m)!} \sin^{2}\theta
  P_m^{\prime}(\cos\theta),
\label{b26}
\end{eqnarray}
and eigennumber values
\begin{equation}
q_m = -m(m+1).
\label{b27}
\end{equation}
Here $P_m$ are Legendre polynomials and prime means their
derivative. As a result, for the full operator $\hat{\cal L}$
we have the following set of eigenfunctions
\begin{enumerate}
\item
$m=1$
\begin{itemize}
\item
$\Phi_1^{(1)} = r^2\sin^2\theta$ -- homogeneous flow,
\item
$\Phi_1^{(2)} = \sin^2\theta/r$ -- dipole flow.
\end{itemize}
\item
$m=2$
\begin{itemize}
\item
$\Phi_2^{(1)} = r^3\sin^2\theta\cos\theta$ -- zero point,
\item
$\Phi_2^{(2)} = \sin^2\theta\cos\theta/r^2$ -- quadrupole flow.
\end{itemize}
\item
\dots
\end{enumerate}

At first glance, this point is absolutely clear so that it is impossible
to encounter here any trouble. Nevertheless, it is not so. Indeed, let us
consider the eigenfunctions corresponding to $m = 0$. The first one is clear:
it is the function
\begin{equation}
\Phi_0^{(1)} = 1-\cos\theta
\label{b28}
\end{equation}
which describes the spherically symmetric accretion/ejection. By the way,
this harmonics alone determines the accretion/ejection rate as for all the
other eigenfunctions with $m > 0$ we have $\Phi_m(r,\pi) = 0$. The uncertainty
is due to the second eigenfunction
\begin{equation}
\Phi_0^{(2)} = r(1-\cos\theta)
\label{b28'}
\end{equation}
for which $\Phi_0(r,\pi) \neq$ const.
It means that this harmonics can be realized only if in the volume
(not in the origin or at infinity) there are the sources or
the sinks of matter.
In all other cases the second eigenfunction for $m = 0$ is to be
dismissed.

\subsection{Examples}

\subsubsection{Bondi-Hoyle Accretion}

As a first example we consider the accretion
onto a moving gravity center (Bondi-Hoyle accretion~\cite{bhoyle}).
This is a classical problem of modern astrophysics. The nature of
Active Galactic Nuclei and Quasars, the nature of jets, the activity of some
galactic X-ray sources are believed to be associated with the accretion
of a gas onto compact objects -- neutron stars and black
holes~\cite{bbr, zn, st, lipunov}.
Nevertheless, only a few exact solutions describing the accretion flow
(even for the simplest adiabatic case) are now known, e.g., the solution
for the spherically symmetric flow we have already discussed.

To construct the nonspherical solution, one may assume that a small
perturbation of a spherically symmetric flow cannot change strongly the
structure of the accretion~\cite{pido1}.  So, it is possible to seek the
solution of the stream equation as a perturbation of the spherically symmetric
solution. First of all, let me recall the main results of the qualitative
theory. It is more convenient to perform the calculations in the reference
frame moving with the gravity center with a velocity $v_{\infty}$. Comparing
the Bondi accretion rate $2\Phi = 4\pi r_{*}^2 n_{*} c_{*}  \sim
(GM)^2n_{\infty}/c_{\infty}^3$ (\ref{a20})--(\ref{a22})
with the flow $\Phi \sim \pi R_{\rm  c}^2 n_{\infty}v_{\infty}$
captured within the capture radius $R_{\rm c}$,
one can evaluate the $R_{\rm c}$ value as
\begin{equation}
R_{\rm c} \sim \varepsilon_1^{-1/2}r_{*},
\label{b29}
\end{equation}
where
\begin{equation}
\varepsilon_1 = \frac{v_{\infty}}{c_{\infty}}.
\label{b30}
\end{equation}
Hence, for $\varepsilon_1 \ll 1$ the capture radius is much larger than
the sonic one, and we can assume that for $r \ll R_{\rm c}$ the flow
structure is similar to the spherically symmetric accretion.
Thus, one can seek the solution of the stream equation (\ref{gs})
in the form
\begin{equation}
\Phi(r,\theta )=\Phi_{0}[1 - \cos\theta + \varepsilon_1 f(r,\theta)].
\label{b31}
\end{equation}
For a nonmoving gravity center we return to the spherically symmetric flow.

As the stream equation (\ref{gs}) now contains $i = 3$ invariants,
so that $b = 2 + 3 - 1 = 4$, it is necessary to specify
four boundary conditions, say
\begin{enumerate}
\item
$v_{\infty} =$ const,
\item
$v_{\varphi} = 0$ (and hence $L = 0$),
\item
$s_{\infty} =$ const,
\item
$E_{\infty} = c_{\infty}^2/(\Gamma+1)$.
\end{enumerate}
In the last relation we neglect the terms $\sim \varepsilon_1^2$.

As a result, the stream equation can be linearized:

\begin{equation}
-\varepsilon_1 D\frac{\partial^{2} f}{\partial r^2}
-\frac{\varepsilon_1}{r^2}(D+1)\sin\theta
\frac{\partial}{\partial\theta}\left(\frac{1}{\sin\theta}
\frac{\partial f}{\partial\theta}\right)
+\varepsilon_1 \left(\frac{2}{r} - \frac{GM}{c_{\rm s}^2r^2}\right)
\frac{\partial f}{\partial r} = 0.
\label{b32}
\end{equation}

This equation has the following properties.
\begin{itemize}
\item
It is linear.
\item
The angular operator coincides with $\hat{\cal L}_{\theta}$ (\ref{b22}).
\item
As all the terms contain a small $\varepsilon_1$ value,
the functions $D$, $c_{\rm s}$, {\it etc.} can be taken
from the zero solution.
\item
As for the spherically symmetric flow the functions $D$, $c_{\rm s}$,
{\it etc.} do not depend on $\theta$, the solution of equation
(\ref{b32}) can be expanded in eigenfunctions of the operator
$\hat{\cal L}_{\theta}$.
\end{itemize}

Thus, the solution can be presented in the form

\begin{equation}
f(r,\theta) = \sum_{m=0}^{\infty} g_m(r) Q_m(\theta),
\label{b33}
\end{equation}
the equations for the radial functions $g_m(r)$ being

\begin{equation}
r^2 D\frac{{\rm d}^{2} g_m}{{\rm d} r^2}
+\left(2r - \frac{GM}{c_{\rm s}^2}\right)
\frac{{\rm d} g_m}{{\rm d} r}
+m(m+1)(D+1)g_m = 0.
\label{b34}
\end{equation}

As to the boundary conditions, they can be formulated as follows:
\begin{enumerate}
\item
No singularity on the sonic surface (where
$r^2N_r = 2r - GM/c_{\rm s}^2 = 0$, $D = 0$)
\begin{equation}
g_m(r_{*}) = 0.
\label{b35}
\end{equation}
\item
A homogeneous flow $\Phi = \pi n_{\infty}v_{\infty} r^2\sin^2\theta$
at infinity, which gives

\begin{equation}
g_1 \rightarrow \frac{1}{2}\,\frac{n_{\infty}c_{\infty}}{n_{*}c_{*}}
\,\frac{r^2}{r_{*}^2}, \qquad g_2, g_3, \dots = 0.
\label{b36}
\end{equation}
\end{enumerate}
As a result, the complete solution can be presented in the form

\begin{equation}
\Phi(r,\theta) = \Phi_{0}[1 - \cos\theta +
\varepsilon_1 g_1(r)\sin^2\theta)],
\label{b37}
\end{equation}
where the radial function $g_1(r)$ is the solution of the
ordinary differential equation (\ref{b34}) for $m = 1$ with
the boundary conditions (\ref{b35}) and (\ref{b36}).

At the present level of personal computers it means that we
succeed in constructing the analytical solution of the problem in
question. It allows us to obtain all the information concerning
the flow structure.
In particular, the sonic surface now has non-spherical form:
\begin{equation}
r_{*}(\theta) = r_{*}\left(1 + 2\varepsilon_1\frac{\Gamma+1}{D}k_2
\cos\theta\right),
\label{b38}
\end{equation}
where the numerical coefficient $k_2 = r_{*} g_1^{\prime}(r_{*})$
can be obtained from the solution (for more details see~\cite{pido1}).
As one can see from Fig. 3, the analytical solution is in full agreement
with the numerical calculations~\cite{hunt} although the parameter
$\varepsilon_1 = 0.6$ is not too small. Finally, the condition
\begin{equation}
g_0 = 0
\end{equation}
(to obtain which an additional consideration is necessary)
shows that the accretion rate is not changed in this approximation.

\begin{figure}[hbtp]
        \centering
        \includegraphics[width=9.truecm]{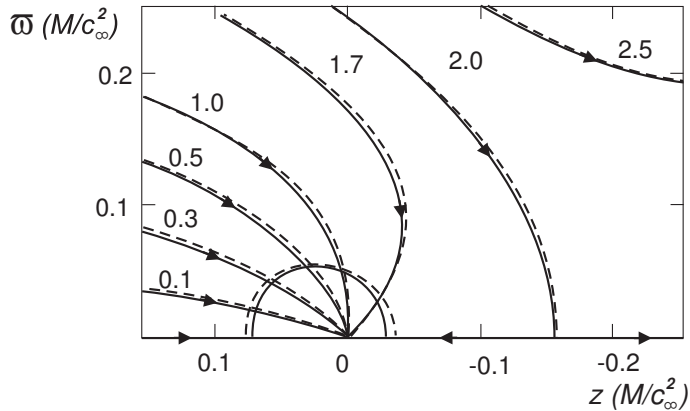}
        \caption[]{Flow structure and shape of the
        sonic surface for $\Gamma = 4/3$,
        $\varepsilon_1 = 0.6$~\cite{pido1}. Labels on the curves
        denote the values of $\Phi/\Phi_0$, and
        dashed curves indicate stream lines and
        the sonic surface obtained numerically
        in~\cite{hunt}}
        \label{fig3}
\end{figure}

In connection with this solution, it is necessary to clarify one point.
As one can easily see, outside the capture radius, our main assumption --
smallness of the disturbance of the spherically symmetric flow --
is not valid. Nevertheless, the constructed solution is correct.
This beautiful property is connected with the already mentioned fact
that for an (approximately) constant concentration $n$ the stream
equation becomes linear. But as we know from the Bondi accretion
(see (\ref{a24})), far from the sonic surface $r\gg r_{*}$
the flow density is actually constant. The same is true for the
homogeneous flow. As a result, for $R_{\rm c} \gg r_{*}$, i.e., for the
case $\varepsilon_1 \ll 1$ under consideration,
in the vicinity of and outside the capture
radius (where the "disturbance" $\sim \varepsilon_1 g_1(r)$ becomes of the
same order as the zero approximation $\sim 1$), the stream
equation is linear. As a result, the sum of two solutions, homogeneous and
spherically symmetric, remains a solution as well.

\subsubsection{Ejection from a Slowly Rotating Star}

Another interesting nonrelativistic example is the transonic ejection from a
slowly rotating star~\cite{tass, cass}. It is necessary to stress from the
very beginning that this example is only illustrative because in reality an
important role is played by the radiation pressure which cannot be included
into consideration within our approach. Nevertheless, the analysis of this
problem helped us to clarify some important features of the Grad-Shafranov
approach~\cite{pido2}.

As a zero approximation we consider the well-known Parker transonic outflow
(\ref{p2})--(\ref{p3}). It means that we assume all the parameters of
the spherically symmetric outflow (the sonic radius $r_{*}$, the velocity
of sound on the sonic surface $c_{*}$, the radial velocity $v_R$ on the
star surface $r = R$, {\it etc}.) to be known. As before, we will seek the
solution in the form
\begin{equation}
\Phi(r,\theta )=\Phi_{0}[1 - \cos\theta + \varepsilon_2^2 f(r,\theta)],
\label{b39}
\end{equation}
where the small parameter is now
\begin{equation}
\varepsilon_2^2 = \frac{\Omega^{2}R^3}{GM}.
\label{b40}
\end{equation}
Here $\Omega$ is the angular velocity of a star.

The problem under consideration needs all the $i = 3$
invariants. Hence, $b = 2 + 3 - 1 = 4$, and it is necessary to specify
four boundary conditions on the star surface $r = r_R(\theta)$ which now
differs from the sphere
\begin{equation}
r_R(\theta) = R[1+\varepsilon_2^{2}\rho(\theta)].
\label{b41}
\end{equation}
Here we introduce the dimensionless parameter $\rho(\theta)\approx 1$.

Let me stress that at first glance there is a disagreement
as we add one degree of freedom (the toroidal rotation with
$v_{\varphi} \neq 0$) while it needs two extra functions
in comparison with the spherically symmetric outflow.
This question will be clarify below.

It is important that for small $\varepsilon_2$ four boundary conditions
can be determined through real physical parameters on the star surface,
e.g., through two thermodynamic functions (say, $T$ and $n$) and two
components of the velocity (say, $v_r$ and $v_{\varphi}$).
As a result, one can express these boundary conditions through four
dimensionless functions $\tau(\theta)$, $\eta(\theta)$, $\omega(\theta)$,
and $h(\theta)$:
\begin{eqnarray}
T(r_R,\theta) & = & T_{R}[1+\varepsilon_2^{2}\tau(\theta)],
\label{b42} \\
n(r_R,\theta) & = & n_{R}[1+\varepsilon_2^{2}\eta(\theta)],
\label{b43} \\
v_{\varphi}(r_R,\theta) & = &
\varepsilon_2\left(\frac{GM}{R}\right)^{1/2}\omega(\theta)\sin\theta,
\label{b44} \\
v_{r}(r_R,\theta) & = & v_{R}[1+\varepsilon_2^{2}h(\theta)].
\end{eqnarray}
Here $\omega(\theta)$, determined as
\begin{equation}
\Omega(r_R,\theta)=\Omega\omega(\theta),
\label{b45}
\end{equation}
describes the differential rotation of the star surface.

Using now the thermodynamic relation
\begin{equation}
{\rm d}s=\frac{1}{\Gamma-1}\frac{{\rm d}T}{T}-\frac{{\rm d}n}{n},
\label{b46}
\end{equation}
one can obtain for the first invariant $s(\theta)$
\begin{equation}
\delta s(\theta)=
\varepsilon_2^2 \left[\frac{1}{\Gamma-1}\tau(\theta)-\eta(\theta)\right].
\label{b47}
\end{equation}
Accordingly, two other invariants can be determined through the boundary
conditions as well
\begin{eqnarray}
\delta E(\theta) & = & \varepsilon_2^{2}v_{R}^{2}h(\theta)+
\varepsilon_2^{2}\frac{GM}{2R}\omega^{2}(\theta)\sin^{2}\theta+
\frac{\varepsilon_2^{2}\Gamma}{\Gamma-1} \frac{T}{m_{\rm p}} \tau(\theta)+
\delta\varphi_{\rm g},
\label{b48} \\
L^2(\theta) & = &
\varepsilon_2^{2}R^{2}\frac{GM}{R}\omega^{2}(\theta)\sin^{2}\theta.
\label{b49}
\end{eqnarray}
Here we use the following expression for the disturbance of the
gravitational potential on the star surface
\begin{equation}
\delta \varphi_{\rm g}(r_R,\theta) = \varepsilon_2^{2}\frac{GM}{R}\rho(\theta).
\label{b50}
\end{equation}

It is of great importance that the possibility of taking the next step,
i.e., writing down the Grad-Shafranov equation, is connected with the simple
geometry of the zero approximation.
Since in the zero approximation the stream function is
$\Phi = 1-\cos\theta$, i.e., it is the function of $\theta$ only (and as
all the derivatives
${\rm d}E/{\rm d}\Phi$, ${\rm d}L/{\rm d}\Phi$, and ${\rm d}s/{\rm d}\Phi$
as well as $L$ itself vanish for a nonrotating outflow),
one can use the relation ${\rm d}\Phi = \Phi_0\sin\theta{\rm d}\theta$.
It allows us to determine the derivatives
${\rm d}E/{\rm d}\Phi$, ${\rm d}L/{\rm d}\Phi$, and ${\rm d}s/{\rm d}\Phi$
in the whole space and not only on the star surface.

As a result, the stream equation can be written as

\begin{eqnarray}
  &  & -\varepsilon_2^{2} \Phi_{0}D\frac{\partial^{2} f}{\partial r^2}-
\frac{\varepsilon_2^2}{r^2}\Phi_{0}(D+1)\sin\theta\frac{\partial}
{\partial\theta}\left(\frac{1}{\sin\theta}
\frac{\partial f}{\partial\theta}\right)+\varepsilon_2^{2}\Phi_{0}
N_{r}\frac{\partial f}{\partial r}= \nonumber \\
  &  & -4\pi^{2}n^{2}r^{2}\sin^{2}\theta(D+1)\frac{{\rm d}E}{{\rm d}\Phi}
+4\pi^{2}n^{2}(D+1)L\frac{{\rm d}L}{{\rm d}\Phi}
\label{gs2} \\
  &  & - 4\pi^{2}n^{2}\frac{\cos\theta}
{\Phi_{0}\sin^{2}\theta}L^2
+ 4\pi^{2}n^{2}r^{2}\sin^{2}\theta\left[(D+1)\frac{T}{m_{\rm p}}+
\frac{\Gamma-1}{\Gamma}c_{s}^2\right]\frac{{\rm d}s}{{\rm d}\Phi}.
\nonumber
\end{eqnarray}
Here $N_r =2/r - 4\pi^2n^2r^2GM/\Phi_0^2$.

The properties of this equation are the same as before.
\begin{itemize}
\item
It is linear.
\item
The angular operator coincides with $\hat{\cal L}_{\theta}$ (\ref{b22}).
\item
As all the terms contain a small $\varepsilon_2^2$ value,
the functions $D$, $c_{\rm s}$, $n$, {\it etc.} can be taken
from the zero approximation.
\item
As for the spherically symmetric flow the functions $D$, $c_{\rm s}$,
$n$, {\it etc.} do not depend on $\theta$, the solution of equation
(\ref{gs2}) can be expanded in eigenfunctions of the operator
$\hat{\cal L}_{\theta}$.
\end{itemize}

Hence, we can again seek the solution in the form

\begin{equation}
f(r,\theta) = \sum_{m=0}^{\infty} g_{m}(r)Q_{m}(\theta).
\label{b51}
\end{equation}
Introducing the dimensionless variables

\begin{equation}
x=\frac{r}{r_{*}}, \quad u=\frac{n}{n_{*}}, \quad a=\frac{c_{s}^2}{c_{*}^2},
\label{b52}
\end{equation}
one can write down the following ordinary differential equations describing
the radial functions $g_m(r)$

\begin{eqnarray}
  &  & (1-x^{4}au^{2})\frac{{\rm d}^{2}g_m}{{\rm d}x^2}+
2(\frac{1}{x}-x^{2}u^{2})\frac{{\rm d}g_m}{{\rm d}x}+
m(m+1)x^{2}au^{2}g_m= \nonumber \\
  &  & \kappa_{m}\frac{R^2}{r_{*}^2}x^{4}au^{4}-
\lambda_{m}\frac{R^2}{r_{*}^2}u^2-
\sigma_{m}x^{6}au^4
\label{gs2'} \\
  & & +\frac{1}{\Gamma}\nu_{m}x^{6}a^{2}u^{4}+
\frac{\Gamma-1}{\Gamma}\nu_{m}x^{2}au^2,
\nonumber
\end{eqnarray}
where $\kappa_{m}$, $\lambda_{m}$, $\sigma_{m}$, and $\nu_{m}$
values are defined as the expansion coefficients
\begin{eqnarray}
\sin\theta\frac{{\rm d}E}{{\rm d}\theta} & = &
\varepsilon_2^{2}c_{*}^2\sum_{m=0}^{\infty} \sigma_{m}Q_{m}(\theta),
\label{coef1} \\
\frac{\cos\theta}{\sin^{2}\theta}L^2 & = & \varepsilon_2^{2}c_{*}^{2}r_{*}^2
\sum_{m=0}^{\infty}\lambda_{m}Q_{m}(\theta), \\
\frac{L}{\sin\theta}\frac{{\rm d}L}{{\rm d}\theta} & = &
\varepsilon_2^{2}c_{*}^{2}r_{*}^2
\sum_{m=0}^{\infty}\kappa_{m}Q_{m}(\theta), \\
\sin\theta\frac{{\rm d}s}{{\rm d}\theta} & = &
\varepsilon_2^{2}\sum_{m=0}^{\infty}\nu_{m} Q_{m}(\theta).
\label{coef4}
\end{eqnarray}

Finally, the functions $a(x)$ and $u(x)$ corresponding to the spherically
symmetric flow for the polytropic equation of state (\ref{a6}) are related
as $a=u^{\Gamma-1}$ and the function $u(x)$ due to (\ref{b18})--(\ref{b20})
can be found from the ordinary differential equation
\begin{equation}
\frac{{\rm d}u}{{\rm d}x}=-2\frac{u}{x}\cdot\frac{1-x^{3}u^2}{1-x^{4}au^{2}}
\label{b53}
\end{equation}
with the boundary conditions (cf. (\ref{a23}))
\begin{equation}
u(1) = 1, \qquad
\left(\frac{{\rm d}u}{{\rm d}x}\right)_{x=1} =
-\frac{4+\sqrt{10-6\Gamma}}{\Gamma+1}.
\label{b55}
\end{equation}

As to the boundary condition to the set of equations (\ref{gs2'}),
they are quite similar to the Bondi-Hoyle accretion (for more details
see~\cite{pido2}):
\begin{itemize}
\item
Condition on the star surface. As
\begin{equation}
{\rm d}\Phi = 2\pi r^{2}nv_{r}\sin\theta {\rm d}\theta =
2\pi R^{2}n_{R}v_{R}[1+\varepsilon_2^{2}
(\eta+h+2\rho)]\sin\theta {\rm d}\theta,
\end{equation}
we have
\begin{equation}
g_{m}(R/r_{*}) = \frac{(2m)!}{2^{m}(m+1)!m!}(\eta_{m}+h_{m}+2\rho_{m}).
\label{b57}
\end{equation}
Here $\eta_m$, $h_m$, and $\rho_m$ are expansion coefficients
in Legendre polynomials, e.g.,
$\eta(\theta)=\sum_{m}\eta_{m}P_{m}(\cos\theta)$.
\item
The absence of singularity on the sonic surface $N_{\theta} = 0$.
This condition gives
\begin{equation}
\varepsilon_{2}^{2}g_{m}(1) = \frac{(2m)!}{2^{m}(m+1)!m!}
\left[\frac{(\delta E)_m}{c_{*}^2} - (\delta s)_m
- \frac{(\delta L/\sin^2\theta)_m}{2c_{*}^2r_{*}^2}\right],
\label{b58}
\end{equation}
where again $(\dots)_m$ means the expansion in Legendre polynomials
which can be found from expressions (\ref{b47})--(\ref{b49}).
\end{itemize}
As a result, equations (\ref{gs2'}) taken together with the
boundary conditions (\ref{b57}) and (\ref{b58})
allow solution of the direct problem, i.e., determination of the flow
structure from the physical boundary condition on the star surface.

Here it is necessary to stress two important points:
\begin{enumerate}
\item
The reason we succeeded in writing the regularity condition
$N_{\theta} = 0$ on the sonic surface (and hence to solve the direct
problem) is again a very simple geometry of the zero approximation.
In particular, in the problem in question the position of the sonic
surface can be taken from the spherically symmetric solution. In the general
case it is not so and it is impossible to write down the condition
$N_{\theta} = 0$ using the known functions $(\delta E)_m$, $(\delta s)_m$,
{\it etc.} on the star surface. The position of the sonic surface
is unknown and is to be found from the solution.
\item
We can now understand the nature of the "extra" boundary condition.
The point is that for $m = 0$ one can take, as was already stressed,
one fundamental solution only, namely, $g_0 = $ const. Another fundamental
solution is unphysical. Hence, there is an additional relation
\begin{equation}
g_0(R) = g_0(r_{*}).
\label{b59}
\end{equation}
This relation determines $h_0$ which, as we see, is not a free parameter.
In other words, we are not fully free in the boundary condition $h(\theta)$:
its zero harmonics is to be found from relation (\ref{b59}). But as
was demonstrated earlier, it is $g_0$ that determines the ejection rate.
Hence, the ejection rate is a function of three parameters only, namely,
zero harmonics of two thermodinamic functions $\eta_0$ and $\tau_0$,
and $v_{\varphi}$. For a spherically symmetric flow $v_{\varphi} = 0$,
and we return to two functions which determine the ejection rate.
As to higher harmonics with $m > 0$, they are free,
and to determine them it is necessary to know four functions on the star
surface. Thus, the spherically symmetric case is degenerate
and it is necessary to be very careful when extending its properties on the
2D flows.
\end{enumerate}

At the end of this section let me formulate some results which can be obtained
under the following simplified assumptions:
\begin{itemize}
\item
Almost the whole of the star mass is in its center,
i.e., $\varphi_{\rm g} = -GM/r$.
\item
No differential rotation, i.e., $\omega(\theta) = 1$.
\item
Von Zeipel law: $T(R,\theta) \propto g_{\rm eff}^{1/4}$,
where $\varphi_{\rm eff} = \varphi_{\rm g} + L^2/r^2$.
\item
No meridional convection, i.e., $v_{\theta}(r_R,\theta) = 0$
(it means that we specify here $v_{\theta}(r_R,\theta)$
instead of $v_r(r_R,\theta)$ ,
i.e., the coefficients $h_0$, $h_1$, {\it etc}. are to be found
from the solution).
\end{itemize}

\vspace{.5cm}

{\bf exercises}

{\it
\begin{enumerate}
\item
Find that the disturbances of the star radius $\rho(\theta)$ in
(\ref{b41}) and the temperature $\tau(\theta)$ in (\ref{b42})
have the form
\begin{equation}
\rho(\theta) = \frac{1}{2}\sin^{2}\theta, \qquad
\tau(\theta) =  -\frac{1}{2}\sin^{2}\theta.
\end{equation}
\item
Show that the only nonzero terms in the expansion (\ref{b51})
correspond to $m = 0$ and $m = 2$, the expansion coefficients
in (\ref{coef1})--(\ref{coef4}) being
\begin{eqnarray}
\sigma_{2} & = & 2\frac{r_{*}}{R}-\frac{5-3\Gamma}{2(\Gamma-1)}
+\frac{1}{2}\frac{v_{R}^2}{c_{*}^2}
-3\frac{v_{R}^2}{c_{*}^2}h_{2}, \\
\lambda_{2} & = & 2\frac{R}{r_{*}}, \quad
\kappa_{2} =  4\frac{R}{r_{*}},  \quad
\nu_{2} =  -\frac{\Gamma}{\Gamma-1},
\end{eqnarray}
and $\sigma_0, ..., \nu_0 = 0 $.
Remember that $h(\theta) = h_{0}+h_{1}\cos\theta+h_{2}P_{2}(\cos\theta)+...$.
\end{enumerate}

}

\vspace{.5cm}

As a result, solving the stream equation (\ref{gs2'}) for $m = 2$, one can
find that
\begin{enumerate}
\item
The ejection rate can be presented as
\begin{equation}
\Phi_{\rm tot} =
2\Phi_{0}\left[1+\frac{\Omega^{2}R^3}{GM}(1+h_{0})\right].
\end{equation}
Here $h_0$ can be obtained from relation (\ref{b59}) (see Table 1)
\begin{equation}
h_{0} = -\frac{1}{6}+\frac{2}{3}\frac{r_{*}/R-R/r_{*}}
{1-v_{R}^2/c_{*}^2}.
\end{equation}
As we see, the rotation increases the ejection rate.

\vspace{.1cm}

{\bf Table 1}

\vspace{.2cm}

\begin{tabular}{|c|c|r|r|r|r|}
\hline
model&$1+h_0$&$h_2$&$q_2$&$b_0$&$b_2$ \\
\hline
$r_{*}/R=1.1$, $\Gamma=4/3$
&  2.9 & $-0.8$ &$-0.40$ & 2.2 &$-0.41$  \\
\hline
$r_{*}/R=2.0$, $\Gamma=4/3$
&  3.2 & $-3.3$ &$-0.71$ & 2.5 &$-0.47$  \\
\hline
$r_{*}/R=10$, $\Gamma=4/3$
&  8.0 &$-56.0$ &$-2.18$ & 5.7 &$-0.92$  \\
\hline
$r_{*}/R=1.1$, $\Gamma=1.1$
&  1.7 & $-0.8$ &$-0.15$ & 1.8 &$-0.37$  \\
\hline
$r_{*}/R=2.0$, $\Gamma=1.1$
&  2.1 & $-2.3$ &$-0.18$ & 2.2 &$-0.40$  \\
\hline
$r_{*}/R=10$, $\Gamma=1.1$
&  7.4 &$-26.0$ &$-0.40$ & 7.2 &$-0.58$   \\
\hline
\end{tabular}

\vspace{.1cm}

\item
Far from the sonic surface $r \gg r_{*}$ the
stream function has the form
\begin{eqnarray}
\lim_{r\rightarrow\infty}
\frac{\Phi(r,\theta)}{\Phi_0} =
(1-\cos\theta)+\frac{\Omega^{2}R^3}{GM}(1+h_0)(1-\cos\theta)
\nonumber \\
+\frac{\Omega^{2}R^3}{GM}q_{2}\sin^{2}\theta\cos\theta,
\label{far}
\end{eqnarray}
where the coefficient $q_{2}$ is tabulated in Table 1 as well.
As there is no $r$-dependence, the outflow becomes purely radial
at large distances.
\item
Accordingly, the asymptotic expression for the concentration $n$
has the form
\begin{equation}
\lim_{r\rightarrow\infty}
\frac{n(r,\theta)}{n_{*}} = \frac{c_{*}}{v_{\infty}}\frac{r_{*}^2}{r^2}
\left[1+\frac{\Omega^{2}R^3}{GM}b_{0}
+\frac{1}{2}\frac{\Omega^{2}R^3}{GM}b_{2}(3\cos^{2}\theta-1)\right],
\end{equation}
where $v^2_{\infty} = v_R^2 - 2GM/R$.
As one can see from Table 1, $b_2 < 0$.
It means that the rotation results in the appearance of a dense disk
in the equatorial plane; this result is well-known~\cite{cass}, but
previously it was obtained by numerical calculations only. Negative
$q_2$ values in (\ref{far}) demonstrate that for $\varepsilon_2 \geq 1$
the most mass outflow is concentrated in the vicinity of the
equatorial plane as well.
\end{enumerate}

\section{Axisymmetric Stationary Flow -- General Relativity}

\subsection{Basic Equations}

We shall show how the Grad-Shafranov approach can be applied to the
axisymmetric stationary flows in the vicinity of a rotating black
hole. Remember that the main difficulty of the General Relativity is
the necessity to work with four-dimensional objects. As a result, we
cannot use our three-dimensional intuition in considering the relativistic
processes.

But there is a convenient language -- $3+1$-split -- which allows work
with three-dimensional vectors even in General Relativity~\cite{tm}.
One can find the detailed introduction into this approach
in the book "Black Holes. The Membrane Paradigm" by K.~Thorne,
R.~Price, and D.~Macdonald~\cite{bhmp}. The main idea is that
for stationary metric the proper time $\tau$ and "the time at
infinity" $t$ are in one-to-one correspondence. It allows the
time $t$ to be separated from spatial coordinates $x^i$ ($i = 1,2,3$).
As a result, all the equations can be written in the simple 3D form,
their physical meaning remaining clear. Below I shall give the main
relations of this approach (see \cite{max} as well).

\subsubsection{Kerr Metric}

The Kerr metric is the metric of a rotating black hole. In the
Boyer-Lindquist coordinates $t$, $r$, $\theta$, and $\varphi$
it has the form
\begin{equation}
{\rm d}s^{2}=-\alpha^{2}{\rm d}t^{2}
+g_{ik}({\rm d}x^{i}+\beta^{i}{\rm d}t)({\rm d}x^{k}+\beta^{k}{\rm d}t),
\label{c1}
\end{equation}
where
\begin{equation}
\alpha=\frac{\rho}{\Sigma}\sqrt{\Delta}
\label{c2}
\end{equation}
($\alpha$ is the lapse function or the red shift), and
\begin{eqnarray}
\beta^{r} = \beta^{\theta}=0, \qquad
\beta^{\varphi}=-\omega=-\frac{2aMr}{\Sigma^{2}}
\label{c3}
\end{eqnarray}
($\omega$ is the Lense--Thirring angular velocity;
remember that $\beta^{\varphi}$ is the contravariant component).
Finally, it is important that the 3D metric $g_{ik}$ in (\ref{c1})
be diagonal
\begin{eqnarray}
g_{rr}=\frac{\rho^{2}}{\Delta},\qquad
g_{\theta \theta}=\rho^{2}, \qquad
g_{\varphi \varphi}=\varpi^{2}.
\label{c4}
\end{eqnarray}
Here $M$ and $a$ are respectively the black hole mass and the angular
momentum per unit mass ($a=J/M$), and we introduce the standard notation
\begin{eqnarray}
 & & \Delta=r^{2}+a^{2}-2Mr, \qquad
 \rho^{2}=r^{2}+a^{2}\cos^{2}\theta, \nonumber\\
& & \Sigma^{2}=(r^{2}+a^{2})^{2}-a^{2}\Delta\sin^{2}\theta, \qquad
 \varpi=\frac{\Sigma}{\rho}\sin\theta.
\label{c5}
\end{eqnarray}
Units where $c=G=1$ are used below.

This metric has the following properties.
\begin{itemize}
\item
The Kerr metric is axisymmetric and stationary. It is precisely
what is needed for using the Grad-Shafranov approach.
\item
The Kerr metric is two-parametric, i.e., it depends on two parameters: the mass
$M$ and the rotation parameter $a$.
\item
It transforms to the Schwarzschild metric for a nonrotating black hole:
$g_{rr} = \alpha^{-2},\, g_{\theta\theta} = r^2, \,
g_{\varphi\varphi} = r^2\sin^2\theta$.
Here $\alpha^2 = 1-2M/r$.
\item
In the limit $r \gg 2M$ Boyer-Lindquist coordinates
coincide with the spherical ones: $g_{rr} = 1, \, g_{\theta\theta} = r^2,
\, g_{\varphi\varphi} = r^2\sin^2\theta$.
\item
Lapse function $\alpha$
\begin{itemize}
\item
The lapse function $\alpha$ describes the lapse (the red shift)
between the proper time $\tau$
and the time at infinity $t$: ${\rm d}\tau = \alpha{\rm d}t$.
\item
The condition $\alpha = 0$ determines the position of the horizon
\begin{equation}
r_{\rm g} = M + \sqrt{M^2 - a^2}.
\label{c6}
\end{equation}
\item
Boyer-Lindquist coordinates do not describe the space-time inside the horizon;
for $r = r_{\rm g}$ the metric has a coordinate singularity.
\end{itemize}
\item
Lense-Thirring angular velocity $\omega$
\begin{itemize}
\item
The Lense-Thirring angular velocity $\omega$ corresponds to the proper motion
of the space around a black hole.
\item
By definition, $\Omega_{\rm H} = \omega(r_{\rm g})$
is the black hole angular velocity (does not depend on $\theta$).
\item
$\omega \propto a$, $\Omega_{\rm H}r_{\rm g} = a/(2M)$.
\end{itemize}
\item
Convenient reference frame -- ZAMO
\begin{itemize}
\item
ZAMO (Zero Angular Momentum Observers)~\cite{bhmp}
are located at a constant radius $r =$ const, $\theta =$ const, but
they rotate with the Lense--Thirring angular velocity
${\rm d}\varphi/{\rm d}t=\omega$.
\item
For ZAMO the four-dimensional metric $g_{\alpha\beta}$ is diagonal,
the spatial 3D metric $g_{ik}$ coinciding with (\ref{c4}).
\item
No gyroscope rotation in the local experiment.
\end{itemize}
\end{itemize}

To clarify the physical meaning of the values $\alpha$ and $\omega$,
let us consider the motion of particles in the gravitational field of a
rotating black hole. Then, one can rewrite the four-dimensional equation
of motion
\begin{equation}
\frac{{\rm d}^2 x^{\alpha}}{{\rm d}s^2}
+ \Gamma^{\alpha}_{\beta\gamma}\frac{{\rm d}x^{\beta}}{{\rm d}s}
\,\frac{{\rm d}x^{\gamma}}{{\rm d}s} = 0
\label{c7}
\end{equation}
in the simple 3D form
\begin{equation}
\frac{{\rm d}p_i}{{\rm d}\tau} =
\frac{m_{\rm p}}{\sqrt{1-v^2}}g_i
+ H_{ik}\frac{m_{\rm p}v^k}{\sqrt{1-v^2}},
\label{c8}
\end{equation}
where
\begin{eqnarray}
{\bf g} & = &  -\frac{1}{\alpha}\nabla\alpha,
\label{c9} \\
H_{ik} & = & \frac{1}{\alpha}\nabla_{i}\beta_{k}.
\label{c10}
\end{eqnarray}
Remember that
\begin{itemize}
\item
The Greek indices $\alpha$, $\beta$, and $\gamma$ are four-dimensional,
while the Latin $i$, $j$, and $k$ are three-dimensional.
\item
$\tau$ is the proper time, and all three-dimensional vectors
are measured by ZAMO.
\item
$\nabla_{i}$ means the covariant derivative in the three-dimensional metric
(\ref{c4}).
\end{itemize}

In a weak gravitational field, i.e., far from a black hole
there is very nice analogy between the gravitational and
electrodynamic equations. Indeed, the equation of motion
(\ref{c8}) can be rewritten as
\begin{equation}
m_{\rm p}\frac{{\rm d}^2{\bf r}}{{\rm d}\tau^2} =
m_{\rm p}\left({\bf g} +
\frac{{\rm d}{\bf r}}{{\rm d}\tau} \times {\bf H}\right),
\label{c11}
\end{equation}
where
\begin{equation}
{\bf g} = -\nabla\alpha, \qquad {\bf H} = \nabla \times {\bf \beta},
\label{c12}
\end{equation}
$\alpha$ and ${\bf \beta}$ playing the role of the scalar and vector
potentials, respectively.
Moreover, the Einstein equations in a weak gravitational field
are quite similar to the Maxwell equations
\begin{eqnarray}
\nabla \cdot {\bf g} & = & - 4\pi\rho_{\rm m}, \\
\nabla \times {\bf g} & = & 0, \\
\nabla \cdot {\bf H} & = & 0, \\
\nabla \times {\bf H} & = & -16 \pi \rho_{\rm m} {\bf v}.
\end{eqnarray}
In other words, the gravitational field ${\bf g}$ is analogous to the
electric field while the new (so-called gravitomagnetic) field ${\bf H}$ --
to the magnetic one which is proportional to the angular velocity of a
black hole. The sources of the gravitoelectric field ${\bf g}$ are masses,
and the sources of the gravitomagnetic field ${\bf H}$ are mass currents.

For example, for a rotating sphere with mass $M$ and angular momentum
${\bf J}$ the fields outside the sphere are~\cite{bhmp}
\begin{eqnarray}
{\bf g} & = & -\frac{M}{r^2}{\bf e}_{\hat r}, \\
\label{c13}
{\bf H} & = & 2\frac{{\bf J} - 3{\bf e}_{\hat r}({\bf Je}_{\hat r})}{r^3},
\label{c14}
\end{eqnarray}
i.e., the rotation induces a dipole gravitomagnetic field around the
rotating body. The appearance of an additional
gravitomagnetic force is the most important consequence of the black hole
rotation.

To summarize, the $3+1$ language allows the description of physical
processes in a clear 3D form. If we simultaneously use ZAMO
as a reference frame, no expressions will contain extra terms.
In a sense, ZAMO is an inertial frame, at any case in the $\varphi$
direction. As a result, within the $3+1$-split

\begin{itemize}
\item
All three-dimensional vectors are to be determined from the local
experiment by ZAMO.
\item
All the calculations are to be made in the 3D diagonal metric (\ref{c4}),
e.g.,
\begin{eqnarray}
\nabla \cdot {\bf A} & = & \frac{1}{\sqrt{g}}\frac{\partial}{\partial x^i}
\left(\sqrt{g}A^i\right),
\label{c15} \\
\nabla \times {\bf A}
&= & \frac{1}{\sqrt{g}}
\pmatrix{
\sqrt{g_{rr}}{\bf e}_{\hat r}
&& \sqrt{g_{\theta\theta}}{\bf e}_{\hat \theta}
&& \sqrt{g_{\varphi\varphi}}{\bf e}_{\hat \varphi}
\cr
\partial/\partial r
&& \partial/\partial \theta
&& \partial/\partial\varphi
\cr
\sqrt{g_{rr}} A_{\hat r}
&& \sqrt{g_{\theta\theta}} A_{\hat \theta}
&& \sqrt{g_{\varphi\varphi}} A_{\hat \varphi}\cr}.
\label{c16}
\end{eqnarray}
\item
All the vector relations remain the same as in the flat space, e.g.,
$\nabla \times (\nabla a) = 0$, $\nabla \cdot(\nabla \times {\bf A}) = 0$.
\end{itemize}

\subsubsection{Thermodynamics}

On the other hand, all thermodynamic functions within the $3+1$
approach are determined in the comoving reference frame, which is the only
invariant one. For this reason, one need no think about the
transformation connected with different reference frames. Actually,
there is the only complification: in the relativistic case one should
work with the relativistic enthalpy $\mu$ including the particle
rest mass
\begin{equation}
\mu=\frac{\rho_{\rm m}+P}{n} \approx m_{\rm p}c^2 + m_{\rm p}w + \dots
\label{c17}
\end{equation}
Here $\rho_{\rm m}$ is the internal energy density. For the polytropic equation
of state $P = k(s)n^{\Gamma}$ (\ref{a6}) we have ($c = 1$)
\begin{eqnarray}
\mu & = & m_{\rm p} + \frac{\Gamma}{\Gamma-1}k(s)n^{\Gamma-1},
\label{c18} \\
c_{\rm s}^2 & = & \frac{1}{\mu}\left(\frac{\partial P}{\partial n}\right)_s
= \frac{\Gamma}{\mu}k(s)n^{\Gamma-1}.
\label{c18'}
\end{eqnarray}
Finally, the relativistic energy-momentum tensor has a symmetrical form
\begin{eqnarray}
T^{\alpha\beta} =
\pmatrix{\varepsilon
&&{\bf S}
\cr
{\bf S}
&& T^{ik}
\cr}
= \pmatrix{(\rho_{\rm m} + Pv^2)\gamma^2
&&(\rho_{\rm m} + P)\gamma{\bf u}
\cr
(\rho_{\rm m} + P)\gamma{\bf u}
&&(\rho_{\rm m} + P)u^iu^k +P g^{ik}
\cr}.
\label{c19}
\end{eqnarray}
Remember that $\gamma$ is the Lorentz-factor of a flow measured by ZAMO.

Now, using the relativistic version of the energy-momentum conservation
law $\nabla_{\alpha}T^{\alpha\beta} = 0$, one can obtain~\cite{tm}
\begin{eqnarray}
 & & -\frac{1}{\alpha}({\bf\beta\nabla})\varepsilon = -\frac{1}{\alpha^2}
{\bf\nabla}\cdot(\alpha^2{\bf S})+H_{ik}T^{ik},
\label{c20}\\
 & & \nabla_{k}T^{k}_{i}+\frac{1}{\alpha}S_{\varphi}\frac{\partial\omega}
{\partial x^{i}}+(\varepsilon\delta^{k}_{i}+T^{k}_{i})\frac{1}{\alpha}
\frac{\partial \alpha}{\partial x^{k}}=0.
\label{c21}
\end{eqnarray}
Here the additional terms in the energy (\ref{c20}) and momentum (\ref{c21})
equations are due to the gravitomagnetic forces.

\subsubsection{Stream Function, etc.}

As in the flat space, one can introduce the stream function $\Phi(r,\theta)$
through the poloidal component of the four-velocity of the flow
${\bf u}_{\rm p}$
\begin{equation}
\alpha n {\bf u}_{\rm p} =
\frac{\nabla \Phi \times {\bf e}_{\hat\varphi}}{2\pi\varpi}.
\label{c22}
\end{equation}
It means the following relations for the physical components
\begin{eqnarray}
\alpha n u_{\hat r} = \frac{1}{2\pi\varpi}(\nabla\Phi)_{\hat\theta},
\label{c23} \\
\alpha n u_{\hat \theta} = -\frac{1}{2\pi\varpi}(\nabla\Phi)_{\hat r}.
\label{c24}
\end{eqnarray}

The definition (\ref{c22}) gives the continuity equation in the form
\begin{equation}
\nabla \cdot (\alpha n {\bf u}) = 0.
\label{c25}
\end{equation}
Let me clarify the extra factor $\alpha$ in (\ref{c25}). The point is that
the 3D continuity equation (\ref{c25}) results from the 4D one
\begin{equation}
\nabla_{\beta} N^{\beta}
= \frac{1}{\sqrt{g_{tt}g}}\frac{\partial}{\partial x^{\beta}}
(\sqrt{g_{tt}g} N^{\beta}),
\label{c26}
\end{equation}
where $g_{tt}$ is equal to $\alpha^2$.

As a result, using the definitions (\ref{c19}) and ({\ref{c22}}),
one can rewrite the energy equation (\ref{c20}) and the $\varphi$
component of the momentum equation (\ref{c21}) as
\begin{eqnarray}
 & & {\bf u} \cdot \nabla(\alpha\mu\gamma)
+ \mu u_{\varphi}{\bf u} \cdot \nabla\omega = 0,
\label{c27} \\
 & & {\bf u} \cdot \nabla(\mu u_{\varphi}) = 0.
\label{c28}
\end{eqnarray}
Hence, two integrals of motion can now be present as
\begin{eqnarray}
E(\Phi) & = & \alpha\mu\gamma +\mu\omega\varpi u_{\hat\varphi},
\label{c29} \\
L(\Phi) & = & \mu\varpi u_{\hat\varphi}.
\label{c30}
\end{eqnarray}

\vspace{.5cm}

{\bf exercise}

{\it

Obtain expressions (\ref{c27})--(\ref{c30}).

}

\vspace{.5cm}

Expressions (\ref{c29}) and (\ref{c30}) are the extension
of the nonrelativistic relations (\ref{b2}) and (\ref{b6}) to the
case of a rotating black hole. Indeed, for $\omega = 0$ we have
for Bernoulli integral
\begin{eqnarray}
E = \gamma\mu\alpha \approx
\left(1+\frac{1}{2}\frac{v^2}{c^2}\right)
\left(m_{\rm p}c^2 +m_{\rm p}w\right)
\left(1 - \frac{GM}{c^2r}\right)
\nonumber \\
\approx m_{\rm p}c^2 +m_{\rm p}\left(\frac{v^2}{2} + w + \varphi_{\rm g}\right)
+ \dots
\label{c31}
\end{eqnarray}
As to the third invariant, we again have
\begin{equation}
s = s(\Phi).
\label{c32}
\end{equation}

\subsubsection{Grad-Shafranov equation}

Using now three invariants $E$, $L$, and $s$, one can write
the poloidal component of the relativistic Euler equation~\cite{fn}
\begin{eqnarray}
nu^{b}\nabla_{b}(\mu u_{a})+\nabla_{a}P
- \mu n(u_{\hat \varphi})^{2} \frac{1}{\varpi}\nabla_{a}\varpi
\nonumber \\
+ \frac{1}{\alpha}\mu n\gamma(\varpi u_{\hat\varphi})\nabla_{a}\omega
+ \frac{1}{\alpha}\mu n\gamma^{2}\nabla_{a}\alpha = 0
\label{euler}
\end{eqnarray}
(here the indices $a$ and $b$ are $r$ and $\theta$ only)
as $[{\rm Euler}]_{\rm p} = [{\rm GS}] \nabla\Phi$, where the
stream equation $[{\rm GS}] = 0$ now looks like

\begin{eqnarray}
 & & -{\cal M}^{2}\left[\alpha\varpi^2\nabla_{k}
\left(\frac{1}{\alpha\varpi^{2}}\nabla^{k}\Phi\right)
+\frac{\nabla^{a} \cdot \Phi\nabla^{b} \cdot \Phi\nabla_{a}\nabla_{b}\Phi}
{({\bf \nabla}\Phi)^{2}D}
\right]
\nonumber \\
 & & +\frac{{\cal M}^{2}\nabla'_{k}F \cdot \nabla^{k}\Phi}
{2({\bf\nabla}\Phi)^{2}D}
\label{gsrel} \\
 & & +\frac{32\pi^{4}}{{\cal M}^{2}}\frac{\partial}
{\partial \Phi}\left[\varpi^{2}(E-\omega L)^{2}-\alpha^{2}L^{2}\right]
-16\pi^{3}\alpha^2\varpi^2nT\frac{{\rm d}s}{{\rm d}\Phi} = 0.
\nonumber
\end{eqnarray}
Here
\begin{equation}
F = \frac{64\pi^{4}}{{\cal M}^{4}}\left[\varpi^{2}(E-\omega L)^{2}
-\alpha^{2}L^{2}-\varpi^{2}\alpha^{2}\mu^{2}\right],
\end{equation}
and we introduce the thermodynamic function
\begin{equation}
{\cal M}^2 = \frac{4\pi \mu}{n}.
\end{equation}
Next, the operator $\nabla_{k}'$ acts on all the variables except ${\cal M}^2$,
and $\partial/\partial \Phi$ on the invariants $E(\Phi)$, $L(\Phi)$,
and $s(\Phi)$ only. Finally, now the denominator $D$ is
\begin{equation}
D = -1 + \frac{1}{u^{2}_{\rm p}}\frac{c^{2}_{\rm s}}{1-c^{2}_{\rm s}}.
\label{c33}
\end{equation}
Here the physical component of the poloidal four-velocity $u_{\rm p}$
can be found from (\ref{c22}).
In a compact form the relativistic stream equation looks like

\begin{eqnarray}
 & & -\alpha\varpi^2\nabla_{k}
\left(\frac{{\cal M}^2}{\alpha\varpi^{2}}\nabla^{k}\Phi\right)
+\frac{32\pi^{4}}{{\cal M}^{2}}\frac{\partial}
{\partial \Phi}\left[\varpi^{2}(E-\omega L)^{2}-\alpha^{2}L^{2}\right]
\nonumber \\
 & &
-16\pi^{3}\alpha^2\varpi^2nT\frac{{\rm d}s}{{\rm d}\Phi}=0.
\label{gsrelcomp}
\end{eqnarray}
The hydrodynamical version of the Grad-Shafranov equation in the Kerr
metric was first formulated in~\cite{anderson} using four-dimensional
notation and in~\cite{bpariev} within the $3+1$-split language.

As before, the Grad-Shafranov equation is to be supplemented with the Bernoulli
equation ($\gamma^2 = 1 + u_{\hat \varphi}^2 + u_{\rm p}^2$) which can now be
written as
\begin{equation}
(E-\omega L)^{2}=\alpha^{2}\mu^{2}+\frac{\alpha^2}{\varpi^2}L^2
+\frac{{\cal M}^4}{64\pi^{4}\varpi^2}(\nabla\Phi)^2.
\label{c34}
\end{equation}
Remember that in (\ref{gsrel}) and (\ref{c34})
the relativistic enthalpy $\mu$ is to be considered as a function of
${\cal M}^2$ and $s$: $\mu = \mu({\cal M}^2,s)$. In the general case the
appropriate differential connection is~\cite{bpariev}
\begin{equation}
{\rm d}\mu = -\frac{c_{\rm s}^2}{1-c_{\rm s}^2}
\mu\frac{{\rm d}{\cal M}^2}{{\cal M}^2}
+ \frac{1}{1 - c_{\rm s}^2}
\left[\frac{1}{n}\left(\frac{\partial P}{\partial s}\right)_{n}
+T\right]{\rm d}s.
\label{c35}
\end{equation}
In particular, as in the nonrelativistic case, the Bernoulli equation
(\ref{c34}) determines ${\cal M}^2$ in the implicit form
through the flux $\Phi$ and three integrals of motion:
${\cal M}^2 = {\cal M}^2(\nabla\Phi;E,L,s)$.

Finally, it is convenient to use another form of the poloidal four-velocity
$u_{\rm p}$:

\begin{equation}
u_{\rm p}^2
= \frac{E^2 - \alpha^2 L^2/\varpi^2 -\alpha^2\mu^2}{\alpha^2\mu^2}.
\label{u_p}
\end{equation}
We see that $u_{\rm p} \rightarrow \infty$ as $\alpha^{-1}$ at the horizon.
As was already stressed, it results from our choice of the reference frame
which has a coordinate singularity for $r = r_{\rm g}$.
Relation (\ref{u_p}) suggests a very important conclusion.
The flow in a close vicinity of a black hole is to be supersonic
($u_{\rm p} > c_{\rm s}$).

\subsection{Examples}

\subsubsection{Exact solutions}

\begin{enumerate}
\item
Spherically symmetric accretion

If the flow velocity at infinity is equal to zero ($\gamma_{\infty}=1$)
and thermodynamic conditions are homogeneous, then $E=\mu_{\infty}=$ const
and $s=s_{\infty}=$ const. So, the two thermodynamic functions at infinity
determine two integrals of motion $E$ and $s$. Finally, for a spherically
symmetric flow one can put $L =0$. Under such conditions the stream equation
(\ref{gsrel}) has a trivial solution $\Phi =\Phi_{0}(1-\cos\theta)$,
and the accretion rate $2\Phi_0$ is to be determined from the
regularity conditions on the sonic surface $r=r_{*}$.

As a result, we have the following expression for the sonic radius
\begin{equation}
r_{*}=\frac{M}{2}\cdot\left(\frac{1}{c_{*}^2}+3\right),
\label{c36}
\end{equation}
so that for $c_{*}^2 \ll 1$  we return to the nonrelativistic
expression (\ref{p2}). As to the relation between
$c_{*}^2$ and $c_{\infty}^2$, it can be found
from the condition~\cite{pido2}
\begin{equation}
\frac{1}{\left[1-c_{\infty}^2/(\Gamma-1)\right]^2}=\frac{1-
4c_{*}^2/(1+3c_{*}^2)}{\left[1-c_{*}^2/(\Gamma-1)\right]^2
\cdot\left(1-c_{*}^2\right)}.
\label{c37}
\end{equation}
In the limit $c_{*}^2\ll 1$ we return to the well-known relation (\ref{a20}).
Next, the values ${\cal M}_{*}^2$ and $\mu_{*}$ on the sonic surface are
\begin{eqnarray}
{\cal M}_{*}^2 & = & {\cal M}_{\infty}^2
\left(\frac{c_{\infty}^2}{c_{*}^2}\right)^{1/(\Gamma-1)}
\left(\frac{\Gamma-1-c_{*}^2}{\Gamma-1-c_{\infty}^2}
\right)^{(2-\Gamma)/(\Gamma-1)},
\label{c38} \\
\mu_{*} & = &\mu_{\infty}\frac{\Gamma-1-c_{\infty}^2}{\Gamma-1-c_{*}^2}.
\label{c39}
\end{eqnarray}
Thus, the accretion rate can be written as $2m_{\rm p}\Phi_{0}$, where
we now have $\Phi_0 = 8\pi^{2}r_{*}^{2}Ec_{*}/{\cal M}_{*}^2$.

\vspace{.5cm}

{\bf exercise}

{\it

Show that in General Relativity a transonic flow takes place even for
$\Gamma = 5/3$.

}

\vspace{.5cm}

Finally, in the supersonic region $r \ll r_{*}$ we have
\begin{eqnarray}
\frac{{\cal M}^2}{{\cal M}_{*}^2} & \simeq &
2\left(\frac{r}{r_{*}}\right)^{3/2},
\label{c41} \\
\frac{c_s^2}{c_{*}^2} & \simeq &
\frac{1}{2^{\Gamma-1}}\cdot\left(\frac{r}{r_{*}}
\right)^{-3(\Gamma-1)/2}.
\label{c42}
\end{eqnarray}
In particular, at the horizon
\begin{equation}
c_{\rm s}^2(r_{\rm g}) = \frac{1}{16^{\Gamma-1}}(c_{*})^{5-3\Gamma}.
\label{c43}
\end{equation}
Hence, for $c_{*}^2 \approx c_{\infty}^2 \ll 1$ the velocity of sound
remains small ($c_{\rm s} \ll 1$) up to the horizon.

It is necessary to stress that an accretion onto black holes
differs sufficiently from the nonrelativistic case. The point is that
all the subsonic trajectories taking place for an accretion onto ordinary
stars (see Fig. 1) have unphysical singularity
$v(r \rightarrow r_{\rm g}) = 0$, $n(r \rightarrow r_{\rm g}) = \infty$
on the horizon. In other words, to support the subsonic flow the infinite
gravitational force
in the vicinity of the horizon is to be balanced by the infinite
pressure gradient. Hence, the only physically reasonable regime of the
accretion onto a black hole is the transonic one.

\item
Accretion of a dust ($P = 0$)

For a dust, the flow lines must coincide with trajectories of particles
freely moving in the gravitational field of a Kerr black hole. For the
case of the motion with $L = 0$ ($u_\varphi=0$) and zero kinetic energy
at infinity $\gamma_{\infty} = 1$, such trajectories are "straight
lines" $\theta = $ const for an arbitrary rotation parameter $a$~\cite{fn}.
Moreover, for $P = 0$ the density of the flow lines can be arbitrary as well.
In other words, the arbitrary function
\begin{equation}
\Phi=\Phi(\theta)
\label{00}
\end{equation}
must be a solution to the stream equation. In particular, it means that
the accretion rate is free. It is not surprising, for the flow is supersonic
in the entire space.

\vspace{.5cm}

{\bf exercise}

{\it

Using the Bernoulli equation (\ref{c34}) and the compact form of the
relativistic stream equation (\ref{gsrelcomp}), check that for
$E = \mu =$ const, $L = 0$, and $s = 0$ the arbitrary function
$\Phi(\theta)$ is a solution.

}

\vspace{.5cm}

\item
Accretion of a gas with $c_{\rm s} = 1$~\cite{pst}

As one can see from (\ref{c33}), for $c_{\rm s}= 1$ we have $D^{-1} = 0$.
Hence, for $E = $ const, $L = 0$, and $s = $ const
the stream equation becomes linear
\begin{equation}
\frac{\Delta}{\rho^2}\frac{\partial^2\Phi}{\partial r^2}
+ \frac{1}{\rho^2}\sin\theta\frac{\partial}{\partial \theta}
\left(\frac{1}{\sin\theta}\,\frac{\partial\Phi}{\partial\theta}\right) = 0.
\label{c44}
\end{equation}
As a result, the solution can again be expanded in eigenfunctions
of the operator $\hat{\cal L}_{\theta}$.
For example, for a moving black hole one can obtain~\cite{pst}
\begin{equation}
\Phi = \Phi_0(1-\cos\theta) + \pi n_{\infty}v_{\infty}
(r^2-2r_{\rm g}r)\sin^2\theta.
\label{c45}
\end{equation}
Here the accretion rate $2\Phi_0$ is arbitrary for another reason -- the flow
remains subsonic up to the horizon.
\end{enumerate}

\subsubsection{Bondi-Hoyle Accretion -- Relativistic Version}

First of all, let us consider the relativistic version of the
Bondi-Hoyle accretion, i.e., accretion onto a moving nonrotating
black hole. The small parameter of the problem is again
$\varepsilon_{1}=v_{\infty}/c_{\infty}$. In the relativistic case
the linearized stream equation for the stream function
$\Phi = \Phi_0[1 - \cos\theta + \varepsilon_1 f(r,\theta)]$
can be written as~\cite{pido1}
\begin{equation}
-\varepsilon_1\alpha^{2} D\frac{\partial^{2} f}{\partial r^2}
-\frac{\varepsilon_1}{\rho^2}(D+1)\sin\theta\frac{\partial}{\partial\theta}
\left(\frac{1}{\sin\theta} \frac{\partial f}{\partial\theta}\right)
+\varepsilon_1 \alpha^{2} N_{r}\frac{\partial f}{\partial r}=0,
\label{c47}
\end{equation}
where now
\begin{equation}
N_r = \frac{2}{r}-\frac{\mu^2}{E^2-\alpha^2\mu^2}\frac{M}{r^2}.
\label{c48}
\end{equation}
We see that this equation have the same properties as the nonrelativistic
equation (\ref{b32}), namely
\begin{itemize}
\item
Equation (\ref{c47}) is linear.
\item
The values of $\mu$, $N_r$, and $D$ should be determined from the
unperturbed Schwarzschild metric for a spherically symmetric
flow.
\item
As $\mu=\mu(r)$, $N_r=N_{r}(r)$, $D=D(r)$, $\alpha^2=1-2{\cal{M}}/r$,
and $\rho=r$, one can expand the solution in eigenfunctions of the operator
$\hat{\cal L}_{\theta}$.
\end{itemize}
But equation (\ref{c47}) has another very important property.
According to (\ref{c33}) and (\ref{u_p}),
\begin{equation}
D + 1 = \frac{\alpha^2\mu^2}{E^2-\alpha^2\mu^2}\cdot
\frac{c_{\rm s}^2}{1-c_{\rm s}^2},
\label{c49}
\end{equation}
so the factor $\alpha^2$ is contained in the all addenda of equation
(\ref{c47}). Hence, equation (\ref{c47}) has no singularity at the horizon.
In particular, it means that it is not necessary to specify any boundary
conditions for $r = r_{\rm g}$. It is not surprising because the horizon
corresponds to the supersonic region which cannot affect the
subsonic flow.

As a result, one can seek the solution of the stream equation in the
form (for more details see~\cite{pido1, ufn})
\begin{equation}
\Phi(r,\theta)=\Phi_{0}[1-\cos\theta+\varepsilon_{1}
g_{1}(r)\sin^{2}\theta],
\label{c50}
\end{equation}
the equation for the radial function $g_{1}(r)$ being
\begin{equation}
-D\frac{{\rm d}^{2}g_1}{{\rm d}r^2}+N_r\frac{{\rm d}g_1}{{\rm d}r}
+2\frac{\mu^2}{E^2-\alpha^2 \mu^2}\cdot\frac{c_{\rm s}^2}{1-c_{\rm s}^2}
\cdot\frac{g_1}{r^2}=0.
\label{c51}
\end{equation}
We see that, as in the nonrelativistic case, the accretion rate is not
changed in the first order of $\varepsilon_1$.

As to the boundary conditions, they are
\begin{enumerate}
\item
The regularity condition on the sonic surface.
It gives $g_{1}(r_{*})=0$.
\item
At infinity
\begin{equation}
g_{1}(r)\rightarrow K(\Gamma)\frac{r^2}{r_{*}^2},
\label{c52}
\end{equation}
where (see Table 2)
\begin{equation}
K(\Gamma) = \frac{1}{2}\frac{M_{*}^2}{M_{\infty}^2}\frac{c_{\infty}}{c_{*}}.
\label{c53}
\end{equation}

\end{enumerate}

\vspace{.1cm}

{\bf Table 2}

\vspace{.2cm}

\begin{tabular}{|l|r|r|r|l|l|l|}
\hline
$\Gamma$        &   1.01   &   1.1  &   1.2  &  1.333  &    1.5  &   1.6 \\
\hline
$K(\Gamma)$     &   0.49   &   0.09 &  0.07  &  0.044  &   0.016 &   0.003 \\
\hline
$k_{1}(\Gamma)$ &   3.00   &   0.56 &  0.46  &  0.31   &   0.12  &   0.023 \\
\hline
$K_{\rm in}(\Gamma)$
               &  $-0.74$  & $-0.09$&$-0.03$ &  0.025  &   0.0081&   0.0002 \\
\hline
\end{tabular}

\vspace{.2cm}

Equation ({\ref{c51}) with boundary conditions 1 and 2 determines
the flow structure of the Bondi-Hoyle accretion onto a nonrotating
black hole. In particular, as in the nonrelativistic case, the shape
of the sonic surface has the form
\begin{equation}
r_{*}(\theta) =
r_{*}\left[1+2\varepsilon_{1}\frac{\Gamma+1}{D_1}
k_{1}(\Gamma)\cos\theta\right],
\end{equation}
where again $k_{1}(\Gamma) = g_{1}^{\prime}(r_{*})r_{*}$.
The $k_{1}(\Gamma)$ values are given in Table 2.

But in general, the flow structure remains the same as for the nonrelativistic
accretion (see Fig. 3). Here one can stress only a single interesting
feature. For $r\ll r_{*}$ the radial function
$g_{1}(r)$ has the asymptotics
\begin{equation}
g_{1}(r)\approx K_{\rm in}(\Gamma)\left(\frac{r}{r_{*}}\right)^{-1/2}
\label{c55}
\end{equation}
($K_{\rm in}$ is given in Table 2 as well). Hence, for
$\varepsilon_{1}>(M/r_{*})^{1/2}$
in the vicinity of the black hole (i.e., for
$r<\varepsilon_{1}^{2}K_{\rm in}^{2} r_{*}$)
the linear approximation is violated, so here one should solve
the complete nonlinear equation (\ref{gsrel}). Since the sign
of the coefficient $K_{\rm in}(\Gamma)$ depends on the polytropic
index $\Gamma$, the region of thickening of the flow lines will either
appear on the front side for $\Gamma > 1.27$ or on the rear side
for $\Gamma < 1.27$. However, it takes place in
the supersonic region which does not affect the subsonic flow for
$r>r_{*}$.

\subsubsection{Accretion onto a Slowly Rotating Black Hole}

Let us now consider the accretion of a gas with $L = 0$
(i.e., $\sigma = 2$ and $b = 2 + 2 - 1 = 3$)
onto a slowly rotating black hole for which the small parameter is
\begin{equation}
\varepsilon_3=\frac{a}{M} \ll 1.
\label{c56}
\end{equation}
In this case the metric $g_{ik}$ (\ref{c4}) differs from the Schwarzschild
one by the values  $\sim \varepsilon_{3}^2$. One can assume that
the thermodynamic functions at infinity, $s_{\infty}$ and $\mu_{\infty}$,
remain the same as for spherically symmetric accretion.
Hence, one can again seek the solution of
the stream equation (\ref{gsrel}) in the form
\begin{equation}
\Phi(r,\theta )=\Phi_{0}[1-\cos\theta + \varepsilon_{3}^{2}f(r,\theta)],
\label{c57}
\end{equation}
where $\Phi_0$ is the flux constant corresponding to a spherically symmetrical
flow. Inserting this form into the stream equation (\ref{gsrel}), we have in
the first order of $\varepsilon_{3}^2$
\begin{eqnarray}
 & & -\varepsilon_{3}^{2}\alpha^{2} D\frac{\partial^{2} f}{\partial r^2}
 -\frac{\varepsilon_{3}^2}{\rho^2}(D+1)\sin\theta\frac{\partial}{\partial
\theta}\left(\frac{1}{\sin\theta} \frac{\partial f}{\partial\theta}\right)
 +\varepsilon_{3}^{2} \alpha^{2} N_{r}\frac{\partial f}{\partial r}
 \nonumber \\
  & & =\frac{a^2}{r^4}\left(1-\frac{2M}{r}\right)\cdot
 \left(1-2\frac{\mu^2}{E^2-\alpha^2\mu^2}
 \frac{M}{r}\right)\sin^{2}\theta\cos\theta,
\label{qq}
\end{eqnarray}
where $N_r$  is given by (\ref{c48}).

The properties of equation (\ref{qq}) are quite similar to those of
equation (\ref{c47}). In particular, equation (\ref{qq}) has no
singularity at the horizon. As a result, one can show that the flux
$\Phi(r,\theta)$ can be presented as~\cite{pido1}
\begin{equation}
\Phi(r,\theta) = \Phi_{0}[(1-\cos \theta)
+\varepsilon_{3}^{2}g_0(1-\cos \theta)
+\varepsilon_{3}^{2}g_2(r)\sin^2\theta\cos\theta].
\label{c59}
\end{equation}
Hence, as for the ejection from a slowly rotating star, the flux
$\Phi(r,\theta)$ contains two harmonics, $m = 0$ and
$m=2$, the radial function $g_2(r)$ satisfying equation
\begin{equation}
-D\frac{{\rm d}^{2}g_2}{{\rm d}r^2}
+N_r\frac{{\rm d}g_2}{{\rm d}r}
+6\frac{\mu^2}{E^2-\alpha^2 \mu^2}
\cdot\frac{c_s^2}{1-c_s^2}\cdot\frac{g_2}{r^2}
=\frac{M^{2}}{r^{4}}
\left(1-2\frac{\mu^2}{E^2-\alpha^{2}\mu^2}\frac{M}{r}\right).
\label{c60}
\end{equation}

The regularity condition on the sonic surface $N_{\theta}(r_{*})=0$ gives
\begin{equation}
g_2(r_{*})=-\frac{1}{2}\frac{M^2}{r_{*}^2}\alpha^2(r_{*}).
\nonumber
\label{c61}
\end{equation}
On the other hand, the condition at infinity (which is the
third boundary condition after $s_{\infty}$ and $\mu_{\infty}$)
gives $g_2(r \rightarrow \infty) = 0$. As a result, the radial
function $g_{2}(r)$ for $r \ll r_*$ can be written in the form
\begin{equation}
g_{2}(r) = -G(r)\frac{M^2}{r_{*}^2}\left(
\frac{r}{r_{*}}\right)^{(1-3\Gamma)/2},
\label{c62}
\end{equation}
where $G(r) \sim 1$. Hence, at the horizon
$g_{2}(r_{\rm g})\sim (M/r_{*})^{(5-3\Gamma)/2}$,
so that the disturbance is small
($\varepsilon_{3}^{2}g_{2}(r)\ll 1$) everywhere outside
a black hole. On the other hand, as $g_2(r) < 0$,
the rotation of a black hole results in the concentration
of the flow lines in the equatorial plane.
Finally, an additional consideration demonstrates
that~\cite{pido1}
\begin{equation}
g_0=-2\frac{M^3}{r_{*}^3}.
\label{c64}
\end{equation}
Thus, the rotation of a black hole diminishes the accretion rate.
In reality $c_{\infty}^2\ll 1$ and hence
$M/r_{*} \ll 1$. As a result, the effects of a black hole
rotation in the vicinity of the sonic surface are actually very small.

To summarize the last two sections, one can say that the simple
zero approximation, namely, a spherically symmetric transonic accretion,
allows us to find analytically 2D structure
for very important astrophysical flows. A similar approach was used
for the construction of an analytical solution for
\begin{itemize}
\item
accretion of a gas with small angular momentum $L$ (the small parameter
$\varepsilon_4^2 = (L/Er_{\rm g})^2$) onto a nonrotating black
hole~\cite{anderson, bmal},
\item
accretion of a gas with a nonrelativistic temperature ($c_{\infty} \ll 1$)
and without angular momentum ($L = 0$)
onto an arbitrary rotating black hole~\cite{pariev}.
\end{itemize}
At present they are the only examples where the analytical solution was found.

\subsubsection{Thin Transonic Disk}

As the last example, let us consider the internal 2D structure of
a thin transonic disk. Such a flow can be realized if the intrinsic
angular momentum of a gas accreting onto a black hole is large enough
($\varepsilon_4 \gg 1$)~\cite{st}. Here for
simplicity we consider a nonrotating (Schwarzschild) black hole only
(for more details see~\cite{btchekh}).

According to the standard disk model~\cite{s, ss, nt, omar},
for $\varepsilon_4 \gg 1$ the matter forms a thin balanced disk and
performs a nearly circular motion with keplerian velocity
$v_{\rm K}(r) \approx \left(GM/r\right)^{1/2}$.
The disk is thin provided that the accreting gas temperature is sufficiently
low and the disk thickness is determined by the pressure gradient
\begin{equation}
H \approx r\frac{c_{\rm s}}{v_{\rm K}}.
\label{2}
\end{equation}
Introducing the viscosity parameter $\alpha_{\rm SS} \leq 1$, relating the
stress tensor $t_{\varphi}^r$ and the pressure as
$t^r_{\varphi} = \alpha_{\rm SS}P$~\cite{ss}, one can obtain
\begin{equation}
\frac{v_r}{v_{ \rm K}} \approx
\alpha_{\rm SS}\frac{c_{\rm s}^2}{v_{\rm K}^2}.
\label{3}
\end{equation}
Hence, for $c_{\rm s} \ll v_{\rm K}$ the
radial velocity $v_r$ remains much smaller than both the keplerian
velocity $v_{\rm K}$ and the velocity of sound $c_{\rm s}$.

The General Relativity effects result in two important properties:
\begin{itemize}
\item
The absence of stable circular orbits for $r < r_0 = 3r_{\rm g}$.
\item
The transonic regime of accretion.
\end{itemize}
The first point means that the accreting matter passing a marginally stable
orbit approaches the black hole horizon sufficiently fast, namely, in
the dynamical time $\tau_{\rm d} \sim [v_r(r_0)/c]^{-1/3}r_{\rm g}/c$.
It is important that such a flow is realized in the absence of viscosity.
The second statement results from the fact that according to (\ref{3})
up to the marginally stable orbit the flow is subsonic while at the
horizon the flow is to be supersonic.

Up to now in the majority of works
the procedure of vertical averaging was used, where the vertical
four-velocity $u_{\hat \theta}$ was assumed to be zero~\cite{pbk}.
As a result, the
vertical component of the dynamic force  $nu^{b}\nabla_{b}(\mu u_{a})$
in (\ref{euler}) was postulated to be inimportant up to horizon. For this
reason the disk thickness was determined by the pressure gradient
even in the supersonic region near the black hole~\cite{alp}.
Here I am going to demonstrate that the assumption $u_{\hat \theta} = 0$ is not
correct. As in the Bondi accretion, the dynamic force is to be important in
the vicinity of the sonic surface.

First of all, let us consider the subsonic region in a close vicinity
of the marginally stable orbit $r_0 = 3r_{\rm g}$ where the poloidal
velocity $u_{\rm p}$ is much smaller than that of sound. Then equation
(\ref{gsrel}) can be significantly simplified by neglecting
the terms proportional to $D^{-1} \sim u_{\rm p}^2/c_{\rm s}^2$.
As a result, we have
\begin{eqnarray}
 & & -{\cal M}^{2}\frac{1}{\alpha}\nabla_{k}
\left(\frac{1}{\alpha\varpi^{2}}\nabla^{k}\Phi\right)
\nonumber \\
 & & +\frac{64\pi^{4}}{\alpha^{2}\varpi^{2}{\cal M}^{2}}
\left(\varpi^{2}E\frac{{\rm d}E}{{\rm d}\Phi}
-\alpha^{2}L\frac{{\rm d}L}{{\rm d}\Phi}\right)
-16\pi^{3}nT\frac{{\rm d}s}{{\rm d}\Phi}=0.
\label{main'}
\end{eqnarray}
This equation describing the subsonic flow is elliptical.

To determine the structure of a two-dimensional subsonic flow
($\sigma = 0$, i.e.,  $b = 2 + 3 - 0 = 5$)
one needs to specify five quantities (three velocity components and
two thermodynamic functions) on an arbitrary surface
$r =r_0(\theta)$. Naturally, we choose it as the surface of the last
stable orbit $r_0 = 3r_{\rm g}$ where
$\alpha_0 = \alpha(r_0) = \sqrt{2/3}$,
$u_{\hat\varphi}(r_0) = 1/\sqrt{3}$, and $\gamma_0 = \gamma(r_0) =
\sqrt{4/3}$~\cite{st}. For the sake of simplicity we consider below the
case where the radial velocity is constant on the surface $r = r_0$
and the toroidal velocity is exactly equal to $u_{\hat\varphi}(r_0)$:
\begin{eqnarray}
u_{\hat r}(r_0,\Theta) & = & -u_0,
\label{ur} \\
u_{\hat \Theta}(r_0,\Theta)  & = & \Theta u_0,
\label{utet} \\
u_{\hat\varphi}(r_0,\Theta) & = & 1/\sqrt{3}.
\end{eqnarray}
Here $u_{\hat \Theta} \ll |u_{\hat r}|$ corresponds to the plane flow
at the marginally stable orbit,
and we introduced the new angular variable $\Theta = \pi/2 -\theta$
($\Theta_{\rm disk} \sim c_0$) which is counted off from the equator in the
vertical direction. Next, we suppose that the velocity of sound is also a
constant on the surface $r = r_0$
\begin{equation}
c_{\rm s}(r_0,\Theta) = c_0.
\label{c0}
\end{equation}
For the polytropic equation of state (\ref{a6}) it means that both the
temperature $T_0 =T(r_0)$ and the relativistic enthalpy $\mu_0 = \mu(r_0)$
are also constant on this surface. According to (\ref{3}), one can find
that for a nonrelativistic temperature $c_{\rm s} \ll 1$  the small
parameter of this problem is
\begin{equation}
\varepsilon_5 = \frac{u_0}{c_0} \sim \alpha_{\rm SS}c_0 \ll 1.
\end{equation}
Finally, as the last, fifth boundary condition
it is convenient to specify the entropy $s(\Phi)$.

Introducing the values
$E_0 =\alpha_0\gamma_0 = \sqrt{8/9}$ and $L_0 = u_{\hat\varphi}(r_0)r_0
= \sqrt{3}r_{\rm g}$, one can write the invariants $E(\Phi)$ and $L(\Phi)$
in the following form
\begin{eqnarray}
E(\Phi) & = & \mu_0 E_0 = {\rm const},
\label{d1} \\
L(\Phi) & = & \mu_0 L_0 \cos\Theta_m.
\label{d2}
\end{eqnarray}
Here $\Theta_m = \Theta_m(\Phi)$ is the angle for
which $\Phi(r_0, \Theta_m) = \Phi (r, \Theta)$. In other words, the
function $\Theta_m(r, \Theta)$ has the meaning of a theta angle
on the last stable orbit. This orbit is connected with a given
point ($r$,$\Theta$) by a line of flow $\Phi(r,\Theta) = $ const.
In particular, $\Theta_m(r_0, \Theta) = \Theta$.

First of all, we see that condition $E = $ const (\ref{d1}) allows us
to rewrite equation (\ref{main'}) in a simpler form
\begin{equation}
\frac{\partial^2 \Phi}{\partial r^2}
+\frac{\cos\Theta}{\alpha^2r^2}\frac{\partial}{\partial\Theta}
\left(\frac{1}{\cos\Theta}\frac{\partial\Phi}{\partial
\Theta}\right)
= -4\pi^2n^2\frac{L}{\mu^2}\,\frac{{\rm d}L}{{\rm d}\Phi}
- 4\pi^{2}n^2r^2\cos^2\Theta\frac{T}{\mu}\,\frac{{\rm d}s}{{\rm d}\Phi}.
\label{main}
\end{equation}
Next, as is shown in Appendix A, for $r=r_0$, the r.h.s. of
equation (\ref{main}) describes
the transverse balance of a pressure gradient and an effective
potential, whereas the l.h.s. corresponds to the
dynamic term $({\bf v}\nabla){\bf v}$. At the marginally
stable orbit it is of the order of
$u_0^2/c_0^2$ and may be dropped. It is therefore natural to
choose the entropy $s(\Phi)$ from the condition of a transverse
balance on the surface $r = r_0$
\begin{equation}
r_0^2\cos^2\Theta_m\frac{{\rm d}s}{{\rm d}\Theta_m} =
-\frac{\Gamma}{c_0^2}\,\frac{L}{\mu_0^2}\,\frac{{\rm d}L}{{\rm d}\Theta_m},
\label{sr0}
\end{equation}
where $L(\Theta_m)$ is determined from the boundary
condition (\ref{d2}). Thus, we have
\begin{equation}
s(\Theta_m) = s(0) - \frac{\Gamma}{3c_ 0^2}\ln(\cos\Theta_m).
\label{sp}
\end{equation}
Owing to (\ref{c18'}), one can show that for $c_{\rm s} =$ const relation
(\ref{sp}) corresponds to the standard concentration profile
\begin{equation}
n(r_0, \Theta) \approx n_0
\exp\left(-\frac{\Gamma}{6c_0^2}\Theta^2\right).
\label{np}
\end{equation}

\vspace{.5 cm}

{\bf exercise}

{\it

Using relations (\ref{a5})--(\ref{a6}),
show that the exact expression for $n(r_0, \Theta)$ is
$n(r_0,\Theta) = n_0(\cos\Theta)^{\Gamma/3c_0^2}$.

}

\vspace{.5 cm}

Finally, the definition (\ref{c22}) results in the following connection
between functions $\Phi$ and $\Theta_m$
\begin{equation}
{\rm d}\Phi = 2\pi \alpha_0 r_0^{2}n(r_0, \Theta_m)u_0
\cos\Theta_m {\rm d}\Theta_m.
\label{d3}
\end{equation}
Hence, due to (\ref{d2}), (\ref{np}), and (\ref{d3}),
the invariant $L(\Phi)$ can be directly determined from the boundary
conditions as well.

Equation (\ref{main}) together with the boundary conditions
(\ref{d2}), (\ref{sp}), (\ref{np}), and (\ref{d3})
and relationship (\ref{utet}) specifying the derivative
$\partial\Phi/\partial r$
determines the structure of the inviscid subsonic flow inside
the marginally stable orbit.
For example, for a nonrelativistic temperature $c_{\rm s}
\ll 1$ we obtain
\begin{equation}
u_{\rm p}^2  = u_0^2 +w^2 +
\frac{1}{3}\left(\Theta_m^2-\Theta^2\right) +\frac{2}{\Gamma -
1}(c_0^2 - c_{\rm s}^2) + \dots
\label{uup}
\end{equation}
Here the quantity $w$, where
\begin{equation}
w^2(r) = \frac{E_0^2-\alpha^2L_0^2/r^2-\alpha^2}{\alpha^2}
\approx \frac{1}{6}\, \left(\frac{r_0 - r}{r_0}\right)^3,
\label{w2}
\end{equation}
depending on the radius $r$ only is a poloidal four-velocity of a free
particle having zero poloidal velocity for $r = r_0$.
As we see, $w^2$ increases very slowly when moving away from
the last stable orbit. Therefore, the
contribution of $w^2$ turns out to be negligibly small and the
term may be safely dropped in most cases.

An important conclusion can be drown directly from (\ref{uup}) in
which for the equatorial plane we have $\Theta_m = \Theta = 0$.
Assuming $u_{\rm p} = c_{\rm s} = c_*$ and neglecting $w^2$, we
find that the velocity of sound $c_*$ on the
sonic surface $r = r_*$, $\Theta = 0$
is of the same order of magnitude as the velocity of sound
on the last stable orbit $c_0$
\begin{equation}
c_* \approx \sqrt{\frac{2}{\Gamma+1}}\,c_0.
\label{cc}
\end{equation}
Since the entropy $s$ remains constant along the flow lines, the gas
concentration remains approximately constant ($n_* \sim n_0$) as well.
In other words, in agreement with the Bondi accretion,
the subsonic flow can be considered as incompressible.

On the other hand, since the density remains almost constant and
for $\varepsilon_5 \ll 1$
the radial velocity increases from $u_0$ to $c_* \sim c_0$, i.e.,
changes over several orders of magnitude, the disk thickness $H$
should change in the same proportion owing to the continuity equation
(see Fig. 4)
\begin{equation}
H(r_*) \approx \frac{u_0}{c_0}H(3r_{\rm g}).
\label{compr}
\end{equation}
As a result, a rapid decrease of the disk thickness should be
accompanied by the appearance of the vertical component of
velocity which also should be taken into account in the Euler
equation (\ref{euler}).

\begin{figure}[hbtp]
        \centering
        \includegraphics[width=9.truecm]{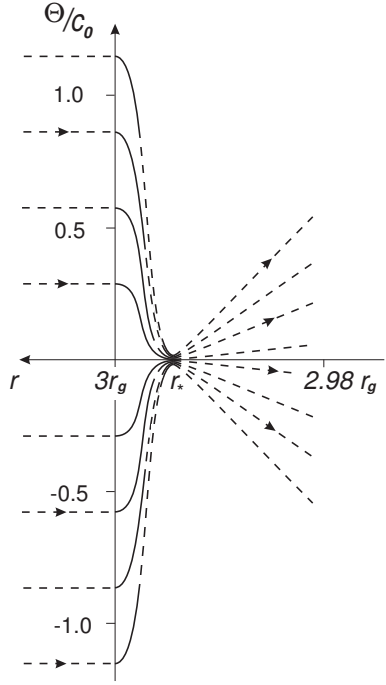}
        \caption[]{The structure of the thin accretion
         disk (actual scale) after passing the marginally
         stable orbit $r = 3r_{\rm g}$ obtained by numerical solving equation
         (\ref{main}) for $c_0 = 10^{-2}$, $u_0 = 10^{-5}$.
         The solid lines correspond to the range of parameters
         $u_{\rm p}^2/c_0^2 < 0.2$, where the solution should not
         differ greatly from the solution of the complete
         equation (\ref{gsrel}). The dashed lines indicate
         an extrapolation of the solution to the sonic-surface
         region. In the vicinity of the sonic surface the flow
         has a form of an ordinary nozzle}
        \label{figure}
\end{figure}

Indeed, as one can find analyzing the asymptotics of equation
(\ref{main})~\cite{btchekh}, in the vicinity of the sonic surface
located at
\begin{equation}
r_*  =  r_0 - \Lambda u_0^{2/3}r_0,
\label{r*}
\end{equation}
where the logarithmic factor
$\Lambda = (3/2)^{2/3}
[\ln(c_0/u_0)]^{2/3} \approx 5 - 7$,
the components of the velocity and the pressure gradient
can be presented as
\begin{eqnarray}
u_{\hat \Theta} & \rightarrow & - \frac{c_0}{u_0}\Theta,
\label{w1} \\
u_{\hat r} & \rightarrow & - c_{*},  \\
- \frac{\nabla_{\hat \theta}P}{\mu} & \rightarrow & \frac{c_0^2}{u_0^2}\,
\frac{\Theta}{r}.
\label{gradp}
\end{eqnarray}
On the other hand, near the sonic surface the radial scale $\delta r$
determining the radial derivatives becomes as small
as the transverse dimension of a disk:
$\delta r \approx H(r_*) \approx u_0r_0$, so that
\begin{equation}
\eta_1 = \frac{r}{n}\,\frac{\partial n}{\partial r} \approx u_0^{-1}.
\label{u00}
\end{equation}
As a result, both components of the dynamic force
\begin{eqnarray}
\frac{u_{\hat \Theta}}{r}\,\frac{\partial u_{\hat
\Theta}}{\partial \Theta} & \rightarrow &
\frac{c_0^2}{u_0^2}\,\frac{\Theta}{r},
\label{w2a}   \\
u_{\hat r}\frac{\partial u_{\hat \Theta}}{\partial r} & \rightarrow &
\frac{c_0^2}{u_0^2}\,\frac{\Theta}{r},
\label{w6}
\end{eqnarray}
become of the order of the pressure gradient (\ref{gradp}).

To check our conclusions one can consider the flow structure in the
vicinity of the sonic surface in more detail. Using again the
expansion theorem (the smooth transonic flow is analytical
at a singular point), one can write down (cf. (\ref{a36})--(\ref{a37}))
\begin{eqnarray}
 n & = & n_*\left(1+\eta_1h+
     \frac{1}{2}\eta_3\Theta^2+\dots\right),
\label{expansionn}\\
 \Theta_m &= & a_0\left(\Theta+a_1h\Theta+\frac{1}{2}a_2h^2\Theta+
            \frac{1}{6}b_0\Theta^3 + \dots
            \right),
\label{expansionthetam}
\end{eqnarray}
where $h = (r-r_*)/r_*$.
Here we assume that all three invariants $E$, $L$, and $s$
are already given, i.e., $i = 0$ and $b = 2 + 0 -1 = 1$.
Hence, as in the planar case, the problem needs one extra
boundary condition.
Now comparing the appropriate coefficients in the Bernoulli (\ref{c34})
and the full stream equation (\ref{gsrelcomp}), one can obtain
neglecting terms $\sim u_0^2/c_0^2$
\begin{eqnarray}
a_0 & = & \left(\frac{2}{\Gamma+1}\right)^%
        {(\Gamma+1)/2(\Gamma-1)}\frac{c_0}{u_0},
\label{a_0}\\
a_1 & = & 2 + \frac{1-\alpha_*^2}{2\alpha_*^2} \approx 2.25,
\label{a_1} \\
a_2 & = & -(\Gamma+1)\eta_1^2,
\label{a_2}\\
b_0 & = &
 \left(\frac{\Gamma + 1}{6}\right) \frac{a_0^2}{c_0^2}, \\
\eta_3 & = & -\frac{2}{3} (\Gamma + 1)\eta_1^2
- \left(\frac{\Gamma-1}{3}\right)\frac{a_0^2}{c_0^2},
  \label{eta3}
\end{eqnarray}
where $\alpha_*^2 = \alpha^2(r_*) \approx 2/3$. In comparison with
the planar case, all the coefficients are expressed here
through the radial logarithmic derivative $\eta_1$ (\ref{u00}).

Let me stress that it is rather difficult to connect the sonic characteristics
$\eta_1 = \eta_1(r_*)$ with physical boundary conditions on the marginally
stable orbit $r = r_0$ (for this it is necessary to know all the expansion
coefficients in (\ref{expansionn}) and (\ref{expansionthetam})).
In particular, it is impossible to formulate the restriction on five
boundary conditions (\ref{ur})--(\ref{c0}) and (\ref{sp}) resulting from
the critical condition on the sonic surface. Nevertheless,
the estimate (\ref{u00}) makes us sure that we know the parameter $\eta_1$
to a high enough accuracy. Then, according to (\ref{a_0})--(\ref{eta3}), all
the other coefficients can be determined exactly.

The coefficients (\ref{a_0})--(\ref{eta3}) have clear physical meaning.
So, $a_0$ gives the compression of flow lines: $a_0 = H(r_0)/H(r_*)$.
In agreement with (\ref{compr}) we have $a_0 \approx c_0/u_0$.
Further, $a_1$ corresponds to the slope of the flow lines with respect to the
equatorial plane. As $a_1 > 0$, in a close vicinity of the sonic
surface the compression of stream lines finishes, so inside the sonic radius
$r < r_*$ the stream lines diverge. On the other hand, as $a_1 \ll u_0^{-1}$,
for $r = r_{*}$ the divergency is still very weak. Hence, in the vicinity
of the sonic surface the flow has a form of an ordinary nozzle
(see Fig. 2).
Finally, as $a_2 \sim \eta_3 \sim b_0 \sim u_0^{-2}$, one can conclude
that the transverse scale of the transonic region $H(r_*)$ does the same as
the longitudinal one. The latter point suggests a very important consequence
that the transonic region is essentially two-dimensional, and so it is
impossible to analyze it within the standard one-dimensional approximation.

Using now the expansions (\ref{expansionn}) and (\ref{expansionthetam}), one
can obtain all the other physical parameters of the transonic flow.
In particular, we have
\begin{eqnarray}
 u_{\rm p}^2 & = & c_*^2
   \left[1-2\eta_1 h+
         \frac{1}{6}(\Gamma - 1) \, \frac{a_0^2}{c_0^2}\Theta^2
                  + \frac{2}{3}(\Gamma + 1)\eta_1^2\Theta^2
   \right],
\nonumber  \\
 c_{\rm s}^2 & = & c_*^2
   \left[1+\left(\Gamma-1\right)\eta_1 h +
        \frac{1}{6}(\Gamma -1 ) \, \frac{a_0^2}{c_0^2}\Theta^2
       - \frac{1}{3}(\Gamma - 1)(\Gamma+1) \eta_1^2
         \Theta^2
   \right].
\nonumber
\end{eqnarray}
As a result, the shape of the sonic surface $u_{\rm p} = c_{\rm s}$
has the standard parabolic form
\begin{equation}
h = \frac{1}{3}\eta_1\Theta^2.
\end{equation}

Thus, the analysis of the hydrodynamic stream equation (\ref{gsrel})
allows us to find a nontrivial structure of the thin transonic disk.
As was shown, the diminishing disk thickness inevitably leads to an
emergence of the vertical velocity component of the accreting matter.
As a result, the dynamic term $({\bf v}\nabla){\bf v}$ in the vertical
balance equation cannot be omitted. In this sense the situation is
completely analogous to the spherically symmetric Bondi accretion
for which the contribution of the dynamic term becomes significant
near the sonic surface and dominant for a supersonic flow.

However, there is an important difference.
For the Bondi accretion the dynamic term $({\bf v}\nabla){\bf v}$
has only one component $v_r\partial v_r/\partial r$ which in the
vicinity of the sonic surface becomes of the same order of magnitude
as both the pressure and the gravity gradients. As to the
thin accretion disk, both components of the dynamic term
$[({\bf v}\nabla){\bf v}]_{\theta}$, (\ref{w2a}) and (\ref{w6}),
become of the same order of magnitude as the pressure gradient,
the role of the effective gravity gradient being unimportant:
$\nabla_{\theta}\varphi_{\rm eff} \sim \Theta/r$,
i.e., it is $c_0^2/u_0^2$ times smaller than the leading terms.
As a result, the structure of a thin transonic disk is quite similar
to an ordinary planar nozzle shown in Fig. 2.
For this reason the critical condition
on the sonic surface does not restrict the accretion rate.

\section{Conclusion}

Thus, in my lecture I have tried to demonstrate the possibilities and
difficulties of the Grad-Shafranov approach. As we have seen, in some
simple cases it does allow us to construct the analytical solutions. In
particular, the Grad-Shafranov approach is very suitable in considering
the (analytical) properties of the flow in the vicinity of the sonic
surface and in determining the number of boundary conditions.

On the other hand, it has been shown that in the general case the
regular procedure does not exist. The point is that the critical
conditions are to be specified on singular surfaces which are not
known from the very beginning and are themselves to be determined
from the solution. Moreover, it is impossible to extend this approach
to nonideal, nonstationary, and non axially symmetric flows. For this
reason it is not surprising that those interested in astrophysics more
than mathematics, passed from the Grad-Shafranov approach to numerical
calculations analyzing absolutely another class of equations, namely,
the time-dependent ones. The only thing one can wish is not to forget
the basic physical results of the Grad-Shafranov approach which remain
true irrespective on the method of calculation.

\Appendix

\section{From Euler to Grad-Shafranov -- the Simplest Way}

Let me show how the nonrelativistic version of the stream equation (\ref{main})
can be directly derived from the Euler equation. For the sake of simplicity
we consider here only a case of nonrelativistic velocities and small
angles $\Theta =\pi/2 - \theta$ in the vicinity of the equatorial plane.

The $\theta$-component of the Euler equation is
\begin{equation}
v_r\frac{\partial v_{\theta}}{\partial r} + v_\theta\frac{\partial
v_{\theta}}{r \partial \theta} + \frac{v_{r}v_{\theta}}{r} -
\frac{v_{\varphi}^2}{r}{\rm cot}\theta =
-\frac{\nabla_{\theta}P}{m_{\rm p}n} - \nabla_{\theta}\varphi_{\rm g}.
\label{A1}
\end{equation}
Adding $v_{\varphi}\partial v_{\varphi}/r\partial \theta$ to both sides
and adding and subtracting $v_{r}\partial v_{r}/r\partial \theta$
in the left-hand side, we get
\begin{equation}
v_r\frac{\partial v_{\theta}}{\partial r} -v_r\frac{\partial
v_r}{r\partial \theta} + \nabla_{\theta}\left(\frac{v^2}{2}\right)
+ \frac{v_{r}v_{\theta}}{r}
= \frac{v_{\varphi}^2}{r}{\rm cot}\theta + v_{\varphi}\frac{\partial
v_{\varphi}}{r\partial \theta} - \frac{\nabla_{\theta}P}{m_{\rm p}n} -
\nabla_{\theta}\varphi_{\rm g}.
\label{A2}
\end{equation}
Using the definition of the Bernoulli integral
$E = v^2/2 + w + \varphi_{\rm g}$
and the thermodynamic relationship ${\rm d}P = m_{\rm p}n {\rm d}w
- n T {\rm d}s$, we get for $E =$ const
\begin{equation}
 v_r\frac{\partial v_{\theta}}{\partial r} -v_r\frac{\partial
 v_r}{r\partial \theta} + \frac{v_{r}v_{\theta}}{r} =
 \frac{rv_{\varphi}\sin\theta}{\varpi^2} \frac{\partial}{r\partial
 \theta} (rv_{\varphi}\sin\theta)
+ \frac{T}{m_{\rm p}}\frac{\partial s}{r\partial\theta}.
\label{A3}
\end{equation}
It follows from the definition (\ref{b1}) that
\begin{equation}
v_r  = \frac{1}{2\pi n
r^2\sin\theta}\frac{\partial\Phi}{\partial\theta}, \qquad
v_{\theta}  = -\frac{1}{2\pi n
r\sin\theta}\frac{\partial\Phi}{\partial r}.
\label{A4}
\end{equation}
Now, assuming $n \approx$ const (this is the case for a
subsonic flow), we obtain
\begin{eqnarray}
-\frac{1}{4\pi^2 n^2}\left(\frac{\partial \Phi}{\partial
\theta}\right) \left[\frac{\partial^2 \Phi}{\partial r^2}
+\frac{\sin\theta}{r^2}\frac{\partial}{\partial\theta}
\left(\frac{1}{\sin\theta}\frac{\partial\Phi}{\partial
\theta}\right) \right]
\nonumber \\
= rv_{\varphi}\sin\theta
\frac{\partial}{\partial \theta}(rv_{\varphi}\sin\theta) +
\varpi^2 \frac{T}{m_{\rm p}}\frac{\partial s}{\partial \theta}.
\label{A5}
\end{eqnarray}
Finally, dividing both sides by $-(\partial \Phi/\partial \theta)$,
we get (\ref{main}).

Hence, whereas the first term in the l.h.s. of (\ref{main}) does
correspond to the component $v_r\partial v_{\theta}/\partial r$, and
the last term in the r.h.s. (for $c_{\rm s} = $ const)
corresponds to the pressure gradient, the role of the term $\propto
L\partial L/\partial\Phi$ proves to be less trivial. It contains
both the effective potential gradient and, in fact, component
$v_\theta\partial v_{\theta}/\partial \theta$. The
former is the leading one near the marginally stable orbit
$r \approx 3r_{\rm g}$, whereas the latter becomes important
only when approaching the sonic surface.

\end{document}